\documentclass[preprint,showpacs,preprintnumbers,amsmath,amssymb]{revtex4}

\usepackage{graphicx}
\usepackage{epsfig}
\usepackage{bm}
\usepackage{amsfonts}
\usepackage{subfigure}
\usepackage{multirow}
\usepackage{color}
\usepackage{float}
\usepackage{xcolor, soul}

\sethlcolor{green}

\newcommand*{\D}{{\rm d}}

\begin{document}



\title{\textbf{Primordial black holes formation in the inflationary model with field-dependent kinetic term for  quartic and natural potentials}}



\author{Milad Solbi\footnote{miladsolbi@gmail.com} and Kayoomars Karami\footnote{kkarami@uok.ac.ir}}
\affiliation{\small{Department of Physics, University of Kurdistan, Pasdaran Street, P.O. Box 66177-15175, Sanandaj, Iran}}

\date{\today}

\begin{abstract}
Within the framework of inflationary model with field-dependent kinetic term for quartic and natural potentials, we investigate generation of the primordial black holes (PBHs) and induced gravitational waves (GWs). In this setup, we consider a kinetic function as $G(\phi)=g_I(\phi)\big(1+g_{II}(\phi)\big)$ and show that in the presence of first term $g_I(\phi)$ both quartic and natural potentials, in contrast to the standard model of inflation, can be consistent, with the 68\% CL of Planck observations. Besides, the second term $g_{II}(\phi)$ can cause a significant enhancement in the primordial curvature perturbations at the small scales which results the PBHs formation. For the both potentials, we obtain an enhancement in the scalar power spectrum at the scales $k\sim10^{12}~\rm Mpc^{-1}$, $10^{8}~\rm Mpc^{-1}$, and $10^{5}~\rm Mpc^{-1}$, which causes PBHs production in mass scales around $10^{-13}M_{\odot}$, $10^{-5}M_{\odot}$, and $10 M_{\odot}$, respectively. Observational constraints confirm that PBHs with a mass scale of $10^{-13}M_{\odot}$ can constitute the total of dark matter in the universe. Furthermore, we estimate the energy density parameter of induced GWs which can be examined by the observation. Also we conclude that it can be parametrized as a power-law function $\Omega_{\rm GW}\sim (f/f_c)^n$, where the power index equals $n=3-2/\ln(f_c/f)$ in the infrared limit $f\ll f_{c}$.
 \end{abstract}



\maketitle

\newpage
\section{Introduction}
In the early universe, primordial curvature perturbations can generate overdense regions. The gravitational collapse of the overdense areas may form primordial black holes (PBHs) after the horizon re-entry \cite{Hawking:1971,Carr:1974}. In contrast to the stellar black holes (BHs), PBHs have a broad mass range, and so they can explain the binary BH detected by the LIGO and Virgo collaborations \cite{Abbott:2016-a,Abbott:2016-b,Abbott:2017-a,Abbott:2017-b,Abbott:2017-c}. The PBHs formed in the early universe can be considered as a candidate for dark matter (DM) \cite{Ivanov:1994,Khlopov1:2005,Frampton:2010,Belotsky:2014,Clesse:2015,Carr:2016,Inomata:2017}. Sasaki et al. \cite{Sasaki:2016} showed that PBHs with a merger rate of around $12-213~ {\rm Gpc^{-3} yr^{-1}}$  and a mass scale of about $ 10 M_{\odot} $ can make up roughly ${\mathcal{O}(1)} \% $ of the total DM. Also, this type of PBHs is well consistent with the LIGO and Virgo observations \cite{Abbott:2017-a}.

Recently, the ultra-short timescale microlensing events were detected in the OGLE data and they prepare an allowed region for PBHs formation. The favored area of OGLE data shows that the abundance of PBHs with a mass scale of ${\mathcal{O}(10^{-5})}M_\odot $ can reach ${\mathcal{O}(10^{-2})}$ \cite{OGLE}. The observations tightly constrain the wide mass range of PBHs as DM, but there is no constraint on the mass scale from ${\mathcal{O}(10^{-13})}M_\odot$ to ${\mathcal{O}(10^{-11})}M_\odot$ \cite{Ali:2017,WD,HSC,EGG,femto,kepler,EROS,CMB-a,CMB-b,Katz:2018,Montero:2019}. Therefore, PBHs in this mass range can demonstrate all the DM in the universe.

The production of PBHs by the collapse of the overdense regions demands an amplitude of the primordial curvature perturbation $A_s$ in order of $\sim{\mathcal{O}(10^{-2})}$ at small scales \cite{sato:2019}. The cosmic microwave background (CMB) anisotropy measurements show that $A_s =2.1 \times 10^{-9}$ at the pivot scale $k_{*}=0.05{\rm Mpc}^{-1}$ \cite{akrami:2018}. It means that $A_s$ must increase nearly seven orders of magnitude at small scales for PBH formation.
A large abundance of PBHs cannot be generated in slow-roll inflationary models \cite{Motohashi:2017,Passaglia:2019}. Therefore, the violation of the slow-roll conditions is needed for PBHs formation. Recent studies have suggested various scenarios for PBHs production \cite{Cai:2018,Ballesteros:2019,Ballesteros:2020a,Ballesteros:2020b,Kamenshchik:2019,Inomata:2018,Ezquiaga:2018,Germani:2017,Di:2018,Ballesteros:2018,Dalianis:2019,
chen:2019,Ozsoy:2018,Atal:2019,mishra:2020,fu:2019,lin:2020,Khlopov:2010,Belotsky1:2014,Belotsky:2019,Braglia:2020,Braglia2:2020,shiPi:2018,Fumagalli:2020a,Sypsas:2020,Dalianis:2020}. For instance, a parametric resonance, which is due to the oscillating sound speed square, can enhance the primordial curvature perturbations \cite{Cai:2018,chen:2019}.
Also, in the single-field model with a non-canonical kinetic term, the curvature perturbations may increase if the sound speed becomes zero \cite{Ballesteros:2019,Kamenshchik:2019}.
Additionally, in the multi-field inflationary models, turning trajectories may lead to the power spectrum enhance at small scales and consequently the PBHs can be produced \cite{Fumagalli:2020a,Braglia:2020,Sypsas:2020}.
One of the  most common approach for PBHs formation is using the inflationary models which have an inflection point. A single field model with an inflection point can lead to the violation of the slow-roll regime \cite{Germani:2017,Di:2018,Ezquiaga:2018,Dalianis:2019}.
Around the inflection point, the inflaton velocity decreases significantly, and the amplitude of the primordial curvature perturbation enhance firmly \cite{fu:2019,lin:2020}. For increasing the primordial curvature perturbation, we require fine-tuning of the model parameters. It should be mentioned that the total number of $e$-fold should remain between 50 and 60 \cite{Passaglia:2019,Sasaki:2018}. Moreover, the models must be consistent with the Planck observations at the CMB scale \cite{akrami:2018}.

Furthermore, the PBHs formation is followed by the generation of the induced gravitational waves (GWs), when the primordial curvature perturbations enhance significantly \cite{Matarrese:1998,Mollerach:2004,Saito:2009,Garcia:2017,Cai:2019-a,Cai:2019-b,Cai:2019-c,Bartolo:2019-a,Bartolo:2019-b,Wang:2019,Fumagalli:2020b,Domenech:2020a,Domenech:2020b,Hajkarim:2019,Kohri:2018,Xu:2020}. In other words, after the horizon re-entry, the collapse of the overdense regions can generate large metric perturbations besides PBHs. In the second-order, the scalar and tensor perturbations may be coupled to each other.  The scalar metric perturbations, through the second-order effect, can generate the stochastic GW background \cite{Cai:2019-a,Cai:2019-b,Cai:2019-c,Bartolo:2019-a,Bartolo:2019-b,Wang:2019,Fumagalli:2020b}. Thus, induced GW signal detection indicates a novel approach to search for PBHs.

The Galileon inflation model is one of the most popular models in the inflation context, which is placed in the subset of Horndeski's theory \cite{Horndeski:1974,kobayashi:2010,Burrage:2010,Tumurtushaa:2019,teimoori:2018}.
In this scenario, the Galileon field acts like a scalar field which is responsible for inflation. In the Minkowski spacetime, the action of this specific scalar field is invariant under the Galilean symmetry $\partial_{\mu}\phi\rightarrow \partial_{\mu}\phi+ b_{\mu}$ \cite{Nicolis:2009,Deffayet:2009a,Deffayet:2009b}. Also, in the Galileon inflation models, the CMB anomalies and the decaying of the CMB power spectrum at the largest scales can be explained when inflaton undergoes the ultra slow-roll phase \cite{Hirano:2016}.
In addition, in \cite{lin:2020} it was shown that in the inflationary model driven by a suitable Galileon term $G(\phi)$, the scalar power spectrum can be enhanced and PBHs are produced. In \cite{lin:2020}, the authors studied the possibility of PBH formation with the Galileon term $G(\phi)=-\frac{1}{2}\frac{d}{\sqrt{\big(\frac{\phi-\phi_c}{c}\big)^2+1}} $ for the power-law potential $V(\phi)=\lambda \phi^{p}$, where $p=2/5$.  

In this paper, our main goal is to investigate the possibility of PBH formation in the framework of inflation with field-dependent kinetic term for quartic and natural potentials. The structure of the paper is as follows. We review the inflation model with field-dependent kinetic term.  in Sec. \ref{sec2}. The mechanism of PBH formation is explained in Sec. \ref{sec3}. In Sec. \ref{sec4}, we study the consequences of reheating to verify whether the primordial curvatures re-enter the horizon during reheating or after. In Sec. \ref{sec5}, we estimate the abundance of PBHs. The induced GWs are investigated in Sec. \ref{sec6}. Finally, Sec. \ref{sec7} is dedicated to our conclusions.

\section{Inflation with field-dependent kinetic term}\label{sec2}
The action of our model is given by \cite{lin:2020}
\begin{equation}
\label{action}
S=\int {\rm d}^4x\sqrt{-g}\left[\frac{M_{\rm pl}^{2}}{2}R+\big(1-2G(\phi) \big)X-V(\phi)\right],
\end{equation}
where $X\equiv -\frac{1}{2}g^{\mu\nu}\phi_{,\mu}\phi_{,\nu}$. In addition,
 $g$ and $R$ are the determinant of the metric $g_{\mu \nu}$ and Ricci scalar, respectively.
Also, $G$ are considered as general functions of the scalar field $\phi$.

From the action (\ref{action}), for a spatially flat Friedmann-Robertson-Walker (FRW) universe the Friedmann equations are obtained as follows \cite{lin:2020,kobayashi:2010,Ohashi:2012}
%
%
\begin{gather}
\label{eom1}
3H^2=\frac{1}{2}\dot{\phi}^2+V(\phi)-\dot{\phi}^2G(\phi),\\
\label{eom2}
2\dot{H}+3H^2+\frac{1}{2}\dot{\phi}^2-V(\phi)-\dot{\phi}^2G(\phi)=0,
\end{gather}
where we take $M_{\rm{pl}}=1/\sqrt{8\pi G}=1$.

Using the action (\ref{action}), the equation of motion governing the scalar field $\phi$ reads
\begin{gather}
\label{eom3}
\ddot{\phi}+3H\dot{\phi}+\frac{V_{,\phi}-\dot{\phi}^2G_{,\phi}}{1-2G(\phi)}=0,
\end{gather}
where $,_{\phi}\equiv {\rm d}/{\rm d}\phi$, and the dot describe the derivative with respect to the cosmic time.

The quadratic action for curvature perturbation ${\cal R}$ at the first order approximation is given by \cite{kobayashi:2010}
\begin{eqnarray}\label{2ndaction}
S^{(2)}=\frac{1}{2} \int\D\tau\D^3x
\tilde{z}^2(1-2G_{,\phi})\left[({\cal R}_\phi')^2-(\Vec{\nabla}{\cal R}_\phi)^2\right],
\end{eqnarray}
where
$\tilde{z}= \frac{a\dot\phi}{H},$
and the prime indicates the derivative with respect to the conformal time $\tau$ \cite{kobayashi:2010}.

In the Fourier space, the Mukhanov-Sasaki (MS) equation can be calculated by varying the action (\ref{2ndaction}) with the respect to the curvature perturbation ${\cal R}$ as follows
\begin{eqnarray}\label{MS_Eq}
u''_k+\left(k^2-\frac{z''}{z}\right)u_k=0,
\end{eqnarray}
where
 $ z=(1-2G)^{1/2}\tilde{z}$, and
$u_k = z{\cal R}_{\phi,k}$ \cite{lin:2020,kobayashi:2010}.
Consequently, the scalar power spectrum can be obtained as
\begin{equation}
{\cal P}_{\cal R}(k)=(2\pi^{2})^{-1}k^{3}\vert u_{k}/z\vert^{2}.
\end{equation}
The slow-roll parameters here are defined as
\begin{equation}
\varepsilon_1\equiv-\frac{\dot{H}}{H^2},\ \varepsilon_2\equiv-\frac{\ddot{\phi}}{H\dot{\phi}},\
 \varepsilon_3\equiv\frac{G_{,\phi}\dot{\phi}^2}{V_{\phi}},
\end{equation}
where $|\varepsilon_i|\ll1$ for $i=1,2,3$.
In the slow-roll conditions, the background Eqs. (\ref{eom1}) and (\ref{eom3}) can be turned to
\begin{gather}
  3H^2\simeq V, \\
  3H\dot{\phi}(1-2G)+V_{\phi}\simeq0.
\end{gather}
Under the slow-roll approximation, the power spectrum of curvature perturbations is given by \cite{lin:2020}
\begin{equation}
\label{eq:ps}
P_{\cal R}=\frac{H^2}{8\pi^2\varepsilon_{1}}\simeq\frac{V^3}{12\pi^2V_{\phi}^2}(1-2G),\\
\end{equation}
and, the scalar spectral index can be estimated as
\begin{equation}
\label{nseq1}
n_s-1=\frac{1}{1-2G}\left(2\eta_V-6\varepsilon_V+
\frac{2G_{\phi}}{1-2G}\sqrt{2\varepsilon_V}\right),
\end{equation}
where $\varepsilon_V\equiv\frac{1}{2}(V'/V)^2$ and $\eta_V\equiv V''/V$. The recent value of the scalar spectral index measured by the Planck satellite is $n_s = 0.9627 \pm 0.0060$ (68$\%$ CL, Planck 2018 TT+lowE) \cite{akrami:2018}.

The tensor power spectrum is given by \cite{lin:2020}
\begin{equation}
\label{ptspec}
P_T=\frac{H^2}{2\pi^2},
\end{equation}
and the tensor-to-scalar ratio can be written as
\begin{equation}
\label{req1}
r\equiv\frac{P_T}{P_{\cal R}}=\frac{16 X(1-2G)}{H^2}.
\end{equation}
There is an upper bound on the tensor-to-scalar ratio provided by the Planck observation as $r< 0.0654$ (68$\%$ CL, Planck 2018 TT+lowE) \cite{akrami:2018}. In the following section, we solve numerically the background equations (\ref{eom2})-(\ref{eom3}) to obtain evolution of the both Hubble parameter and the scalar field. Then, with the help of numerical solution of the MS equation (\ref{MS_Eq}), we estimate the exact value of the scalar power spectrum. Note that in our numerical calculations, we use the slow roll solutions as initial conditions.

%
%

\section{PBH formation mechanism}\label{sec3}
A proper kinetic term can amplify the curvature perturbations at small scales. Also, the proposed function must cause the model to be consistent with the Planck measurements at the pivot scale $k_{*}=0.05{\rm Mpc}^{-1}$. To this aim, we suggest the kinetic function $G(\phi)$ to be parameterized as follows
\begin{equation}\label{g}
G(\phi)=g_I(\phi)\big(1+g_{II}(\phi)\big),
\end{equation}
where
\begin{equation}\label{gI}
g_I(\phi)=-\left(\frac{\phi}{M}\right)^{\alpha},
\end{equation}
\begin{equation}\label{gII}
g_{II}(\phi)=\frac{1}{2}\frac{d}{\sqrt{\big(\frac{\phi-\phi_c}{c}\big)^2+1}}\,.
\end{equation}
In the above, $g_I(\phi)$ is the base kinetic term, and we utilize it to sure that the model can satisfy the observational constraints on $n_s$ and $r$. Also $M$ is a constant with the mass dimension and for $M \rightarrow \infty$, our model recovers the standard inflation.
Note that the term $g_{II}(\phi)$ is responsible for generating the peak in the scalar power spectrum at $\phi=\phi_c $. Also, the value of $g_{II}(\phi)$ is vanishing at distances far from $\phi=\phi_c$. Here, the parameters $\alpha$ and $d$ are dimensionless constants. Also $\phi_c$ and $c$ have the dimension of mass. Additionally, $d$ and $c$ control the height and width of the scalar power spectrum at peak position, respectively.

Note that Lin et al. \cite{lin:2020} has already been studied a similar model like (\ref{g}) in which the selected kinetic term has only one part as $G(\phi)=-\frac{1}{2}\frac{d}{\sqrt{\big(\frac{\phi-\phi_c}{c}\big)^2+1}}$. Also, they considered the power-law potential $V(\phi)=\lambda \phi^{p}$, where $p=2/5$.

\subsection{Quartic inflation with field-dependent kinetic term}
Here, we are interested in investigating the possibility of PBH formation in our setup with quartic potential. The inflationary quartic potential has the following form
\begin{equation}\label{h-potential}
V(\phi)=\frac{\lambda}{4}\phi^{4},
\end{equation}
where $\lambda$ is a constant parameter. At the horizon exit, we set $N_{*}=0$ and take $\phi_{*}\simeq 0.765 $ at the pivot scale, which is estimated by slow-roll approximation. Note that in the framework of standard inflation, i.e. the Einstein gravity, the quartic potential (\ref{h-potential}) is completely rolled out by the Planck 2018 data \cite{akrami:2018}. This motivates us to examine the quartic potential in this scenario to check viability of the model in light of the Planck observations.

In the case of quartic inflation, the model is described by six free parameters ($\alpha$, $\lambda$, $d$, $c$, $\phi_{c}$, $M$). By setting suitable values for $M$ and $\alpha$ one can get the observational parameters $n_s$ and $r$ compatible with the Planck 2018 data. Also, the parameter $\lambda$ is fixed by the amplitude of scalar power spectrum at the CMB scale $k_{*}=0.05~{\rm Mpc}^{-1}$ \cite{akrami:2018}. Thus, $d$, $c$, and $\phi_{c}$ are the remaining free parameters of our model, and they can affect the PBH production.

\begin{table}[H]
  \centering
  \caption{The parameter sets for PBHs production in quartic inflation model. The parameters $\phi_c$ and $c$ are in units of $M_{\rm{pl}}=1$. Here, $M=0.46$ and $\alpha=15$.}
  \begin{tabular}{cccc}
  \hline
  Sets \quad &\quad $\phi_{c}$ \quad & \quad $d$\quad &$c$\quad \\ [0.5ex]
  \hline
  \hline
$\rm{Case~}A_{\rm H}$ \quad &\quad $0.707$ \quad &\quad $2.659\times 10^{8}$ \quad &\quad $2.3\times10^{-11}$ \quad\\[0.5ex]
  \hline
$\rm{Case~}B_{\rm H}$ \quad &\quad $0.739$ \quad &\quad $1.415\times 10^{8}$ \quad &\quad $2.1\times10^{-11}$ \quad\\[0.5ex]
  \hline
$\rm{Case~}C_{\rm H}$ \quad &\quad $0.753$ \quad &\quad $1.265\times 10^{8}$ \quad &\quad $1.8\times10^{-11}$ \quad\\[0.5ex]
  \hline
  \end{tabular}
  \label{tab1}
\end{table}

\begin{table}[H]
  \centering
  \caption{The values of $n_s$, $r$, $k_{\text{peak}}$, ${\cal P}_{\cal R}^\text{peak}$, $M_{\text{PBH}}^{\text{peak}}$ and $f_{\text{PBH}}^{\text{peak}}$ for the cases listed in Table \ref{tab1}.}
  \begin{tabular}{ccccccc}
  \hline
  Sets \quad & \quad $n_{s}$\quad &$r$\quad & \quad$k_{\text{peak}}/\text{\rm Mpc}^{-1}$ \quad &\quad ${\cal P}_{\cal R}^\text{peak}$ \quad & \quad  $M_{\text{PBH}}^{\text{peak}}/M_{\odot}$ \quad & \quad $f_{\text{PBH}}^{\text{peak}}$\\ [0.5ex]
  \hline
  \hline
$\rm{Case~}A_{\rm H}$ \qquad &\quad $0.963$ \quad &\quad $0.057$ \quad &\quad $2.29\times10^{12}$ \quad &\quad $0.035$ \quad &\quad $4.77\times 10^{-13}$ \quad &$0.91$ \\[0.5ex]
  \hline
$\rm{Case~}B_{\rm H}$ \quad &\quad $0.965$ \quad &\quad $0.057$ \quad &\quad $4.52\times10^{8}$ \quad &\quad $0.042$ \quad &\quad $1.82\times10^{-5}$ \quad & $0.031$ \\[0.5ex]
  \hline
$\rm{Case~}C_{\rm H}$ \quad &\quad $0.972$ \quad &\quad $0.056$ \quad &\quad $2.61\times10^{5}$ \quad &\quad $0.052$ \quad &\quad $36.9$ \quad &$0.0017$\\[0.5ex]
  \hline
  \end{tabular}
  \label{tab2}
\end{table}

\begin{figure}[H]
\begin{minipage}[b]{1\textwidth}
\subfigure[\label{fig-phi-h1} ]{ \includegraphics[width=0.45\textwidth]%
{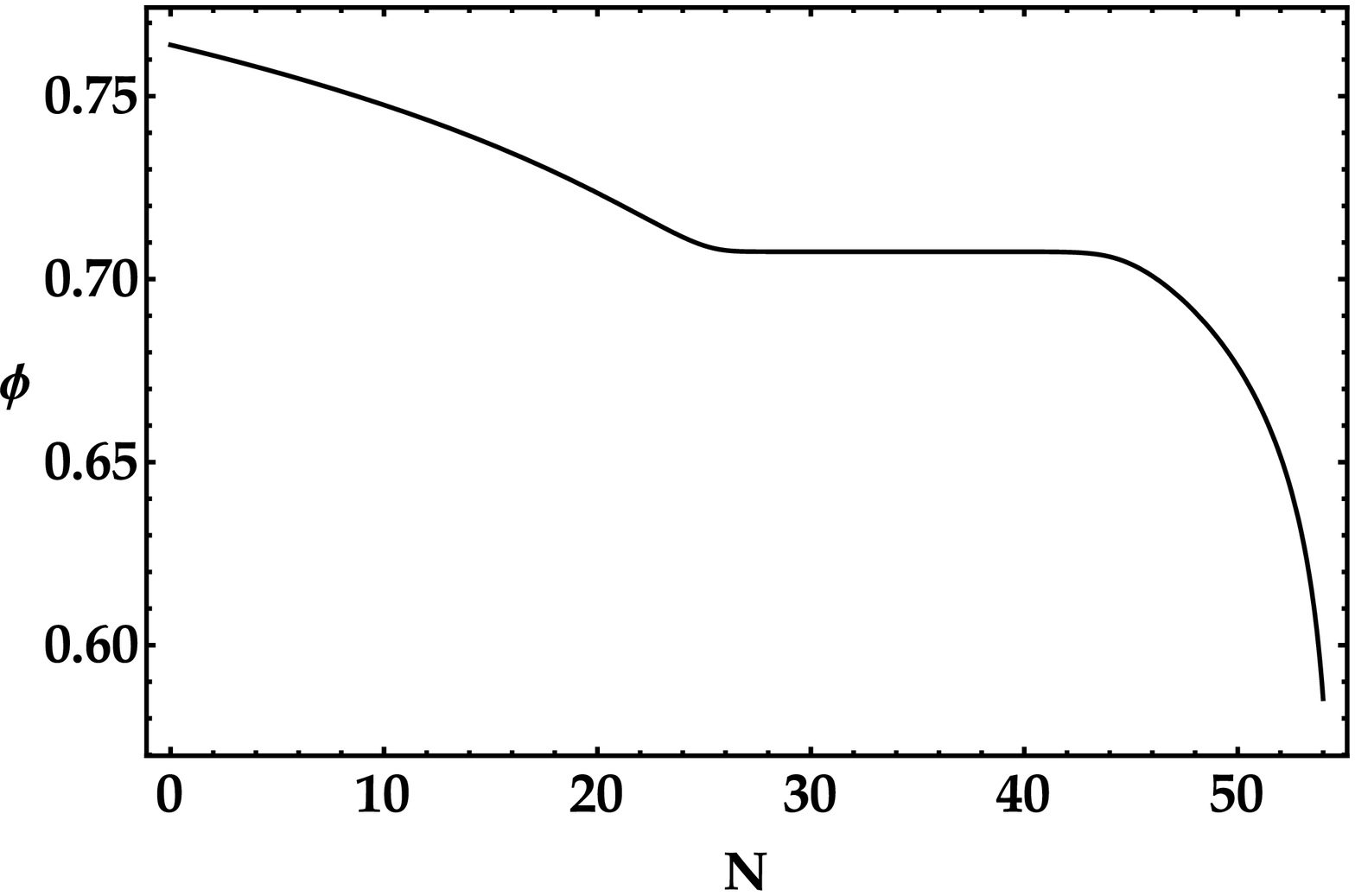}}\hspace{.1cm}
\subfigure[\label{fig-eps-h1}]{ \includegraphics[width=.45\textwidth]%
{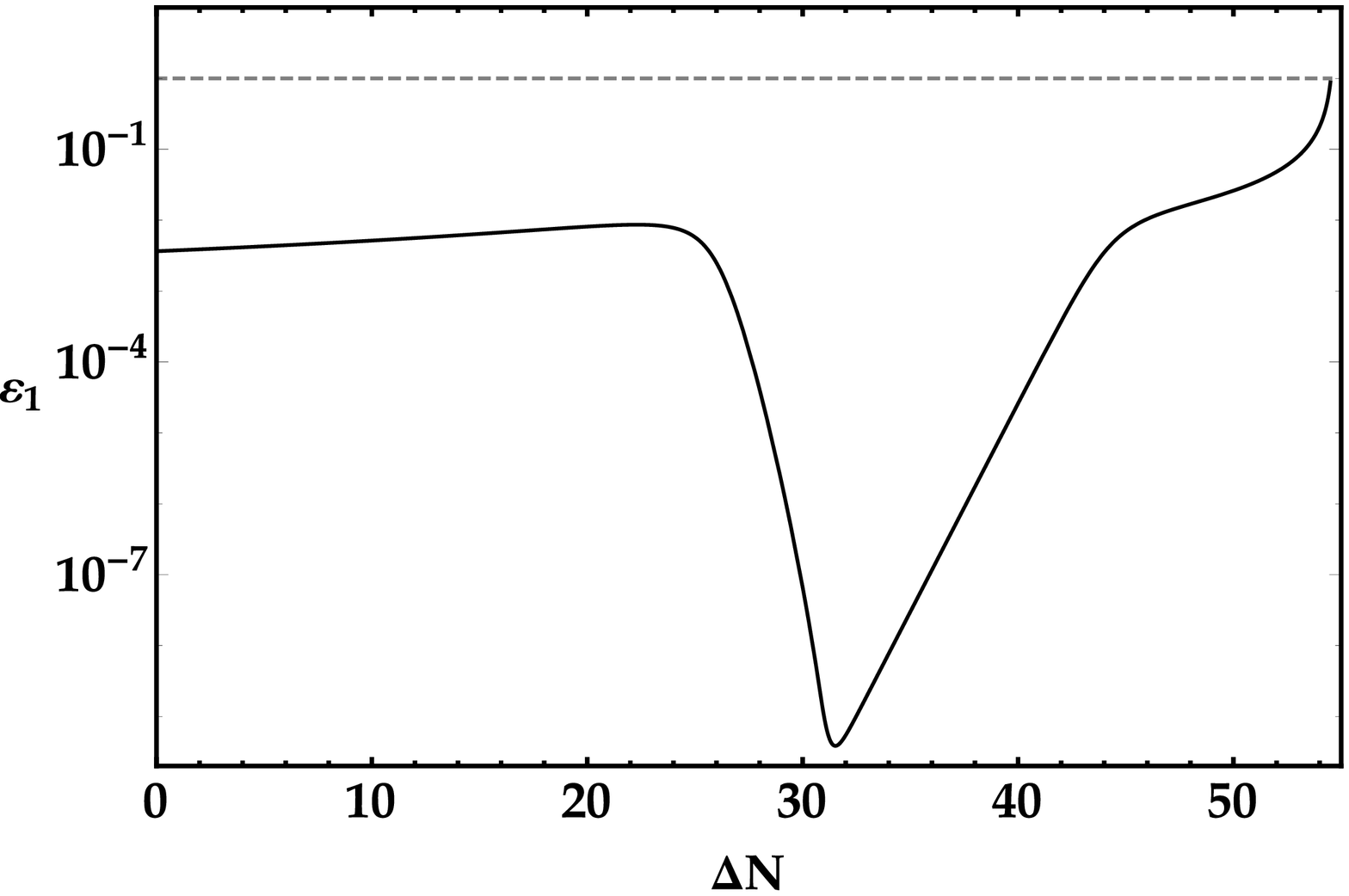}}
\centering
\subfigure[\label{fig-eta-h1}]{
 \includegraphics[width=.45\textwidth]%
{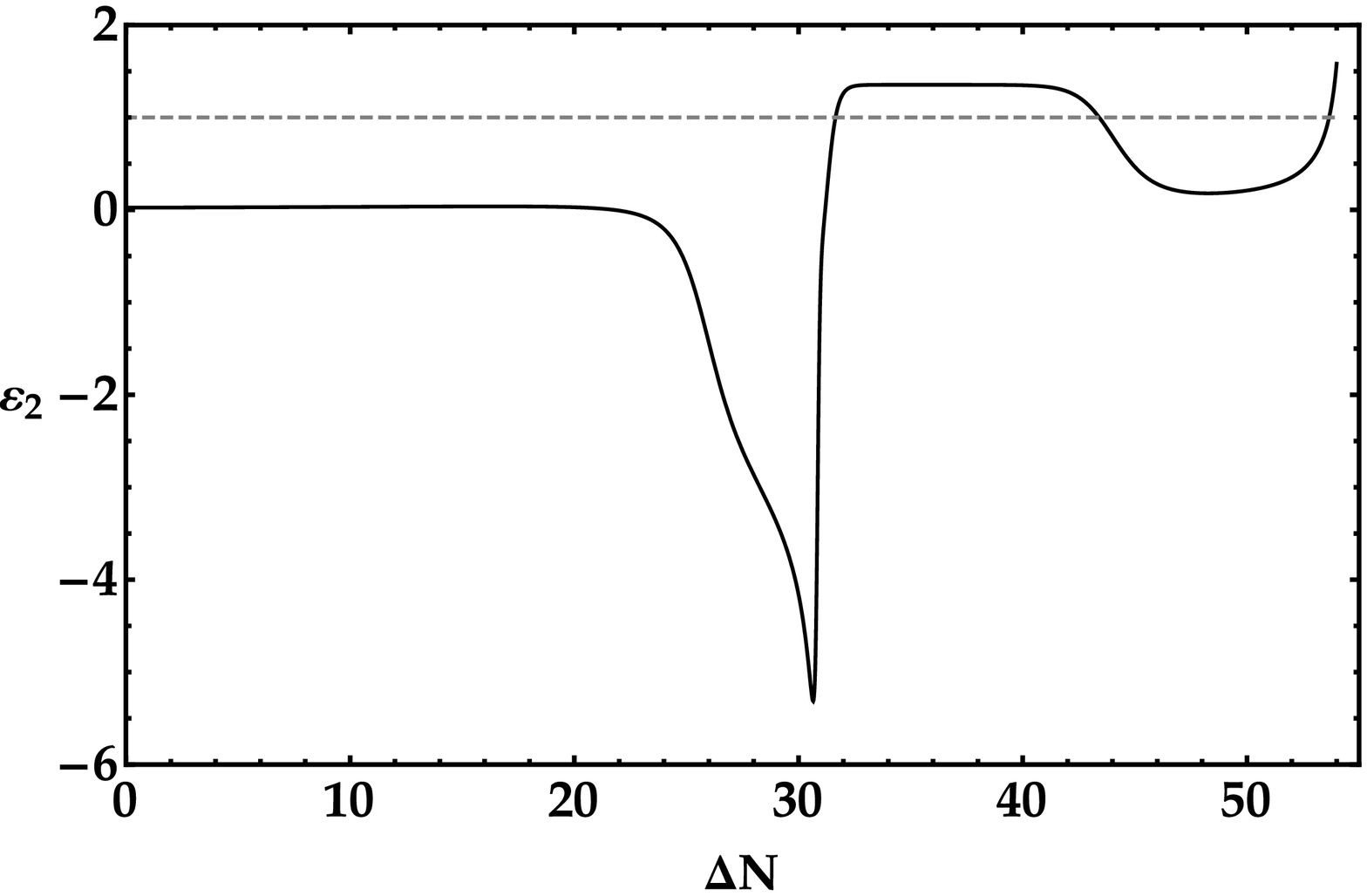}}
\end{minipage}
\caption{Evolutions of (a) the scalar filed $\phi$, (b) the first slow-roll parameter $\varepsilon_1$ and (c) the second slow-roll parameter $\varepsilon_2$ versus the $e$-fold number $N$ for the quartic inflation model. The dashed line shows the violation of the slow-roll condition. The auxiliary parameters are given by the parameter set of case $A_{\rm H}$ in Table \ref{tab1} and we take $N_{*}=0$ at the horizon exit.
  }\label{linear}
\end{figure}

The slow-roll regime fails on the small scales where the peak of the power spectrum appears. It means that, for calculating the exact value of the scalar power spectrum ${\cal P}_{\cal R}(k)=(2\pi^{2})^{-1}k^{3}\vert u_{k}/z\vert^{2}$, we need to solve the MS equation (\ref{MS_Eq}), numerically. We find three sets of parameters, which are listed in Table \ref{tab1}. Here, we set $M=0.46$ and $\alpha=15$ to keep our model predictions consistent with the Planck measurements for $n_s$ and $r$ \cite{akrami:2018}. As shown in Table \ref{tab2}, in the cases $A_H$ and $B_H$ the predictions of the model for $n_s$ and $r$ take place inside the 68$\%$ CL region of the Planck 2018 TT+lowE data \cite{akrami:2018}. In the case $C_H$, the results of $n_s$ and $r$, respectively, are compatible with the 95$\%$ and 68$\%$ CL of the Planck 2018 data \cite{akrami:2018}. Note that these results are in contrast with the result of quartic potential in the standard model of inflation, in which the prediction of the model is completely rolled out in light of the Planck observations \cite{akrami:2018}. The parameter $d$ should be at least in order of ${\cal O}(10^8)$ to the power spectrum peak enhances seven
orders of magnitude at peak position. For fine tuning the parameter $c$ which controls the width of the scalar power spectrum at peak position, we have two restrictions. In one hand, the total number of $e$-folds should remain between 50 and 60. On the other hand, the quantity $g_{II}(\phi)$ should be negligible away from the peak to the usual slow-roll is guaranteed. Using these limitations, we set the parameter $c$ as in order of ${\cal O}(10^{-11})$.
The values of the power spectrum and the corresponding PBHs abundance for these sets are shown in Table \ref{tab2}.
The evolution of the scalar field $\phi$ versus the $e$-fold number $N$ is depicted in Fig. \ref{fig-phi-h1} for the parameter set $A_{\rm H}$.
The flat region in Fig. \ref{fig-phi-h1} is because of the decreasing inflaton velocity in this area where the inflaton undergoes the ultra slow-roll (USR) phase. The value of $\varepsilon_1$ is reduced critically in the USR stage, as shown in the Fig. \ref{fig-eps-h1}. The intense reduction can give a substantial accretion in the scalar power spectrum. In Fig. \ref{fig-eta-h1}, we see that during the USR period, the slow-roll condition $|\varepsilon_2|\ll 1$ is violated.

\begin{figure}[H]
\centering
\includegraphics[scale=0.5]{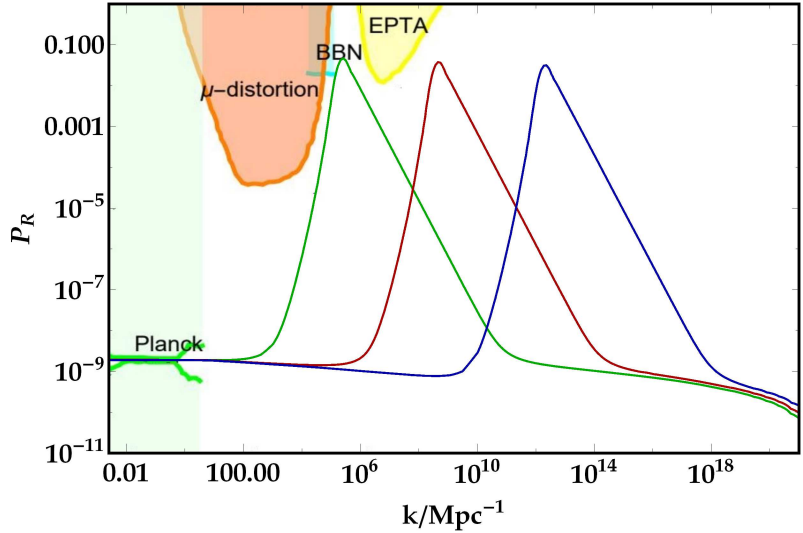}
\vspace{-0.5em}
\caption{The scalar power spectrum ${\cal P}_{\cal R}$ in terms of wavenumber $k$ for the quartic inflation model. The blue, red and green  lines are corresponding to the cases $A_{\rm H}$, $B_{\rm H}$ and $C_{\rm H}$, respectively. The CMB observations exclude the light-green shaded area \cite{akrami:2018}. The orange zone shows the $\mu$-distortion of CMB \cite{Fixsen:1996}. The cyan area represents the effect on the ratio between neutron and proton during the big bang nucleosynthesis (BBN) \cite{Inomata:2016}. The yellow region demonstrates the constraint from the PTA observations \cite{Inomata:2019-a}. }
\label{fig-pr-higgs}
\end{figure}

Figure \ref{fig-pr-higgs} presents the scalar power spectrum for the three parameter sets tabulated in Table \ref{tab1}. As displayed in Fig. \ref{fig-pr-higgs}, the scalar power spectrum grows seven orders from $\sim{\cal O}(10^{-9})$ at the CMB scale to $\sim{\cal O}(10^{-2})$ at the small scales, which is ideally suitable for PBHs formation. Also, the results of our model for the PBHs production are in good agreement with the observational constraints like the observations of CMB $\mu$-distortion, big bang nucleosynthesis (BBN), and pulsar timing array (PTA) \cite{Inomata:2019-a,Inomata:2016,Fixsen:1996}.

\subsection{Natural inflation with field-dependent kinetic term}
In this section, we study the PBHs formation in the framework of inflation with field-dependent kinetic term for the natural potential given by
\begin{equation}\label{potential}
V(\phi)=\lambda^4\left[1+\cos\left(\frac{\phi}{f}\right)\right],
\end{equation}
where $\lambda$ and $f$ are constant. Here we fix the parameter $\lambda$ by using $P_{\cal R}(k_{*})=2.1\times 10^{-9}$ at the CMB scale $k_{*}=0.05~{\rm Mpc}^{-1}$ \cite{akrami:2018}. With the help of slow-roll approximation, we obtain the scalar field at the pivot scale as $\phi_{*}\simeq 0.23 $. Also, we set $N_{*}=0$ at the horizon exit. Note that in the standard inflation setting, the prediction of the  natural potential (\ref{potential}) is not in very good agreement with the latest observations, because its result takes place inside the 95$\%$ CL constraints of Planck 2018 TT+lowE data \cite{akrami:2018}. This motivates us to investigate whether the results of the natural potential in this scenario can be improved in light of the Planck 2018 observations.

For the case of natural potential (\ref{potential}), we have seven free parameters ($\alpha$, $\lambda$, $f$, $d$, $c$, $\phi_{c}$, $M$) for our model.
By choosing the proper values for $M$, $\alpha$, and $f$, one can adjust the values of $n_s$  and $r$ to be compatible with the Planck observations.
Also, the parameter $\lambda$ is fixed by the amplitude of scalar power spectrum at the pivot scale $k_{*}=0.05~{\rm Mpc}^{-1}$ \cite{akrami:2018}. As a result, the free parameters $d$, $c$, and $\phi_{c}$ are only remaining parameters which may affect the generation of PBHs.

For the natural inflation model we obtain the three parameter sets listed in Table \ref{tab3} which can lead to the PBHs generation. By setting $M=1.7$, $\alpha=-1$, and $f=1$, our model is compatible with the Planck observations for $n_s$ and $r$. In the cases $A_{\rm N}$ and $B_{\rm N}$, the results of $n_{s}$ and $r$ are compatible with the Planck observations at the $68\% $ CL  \cite{akrami:2018}. Also in the case $C_{\rm N}$, we get $n_{s}=0.961$ which satisfies the 95$\%$ CL constraint, but the value of $r=0.005$ falls in the $68\%$ CL region of Planck 2018 data \cite{akrami:2018}.

\begin{figure}[H]
\begin{minipage}[b]{1\textwidth}
\subfigure[\label{fig-phi-n1} ]{ \includegraphics[width=0.45\textwidth]%
{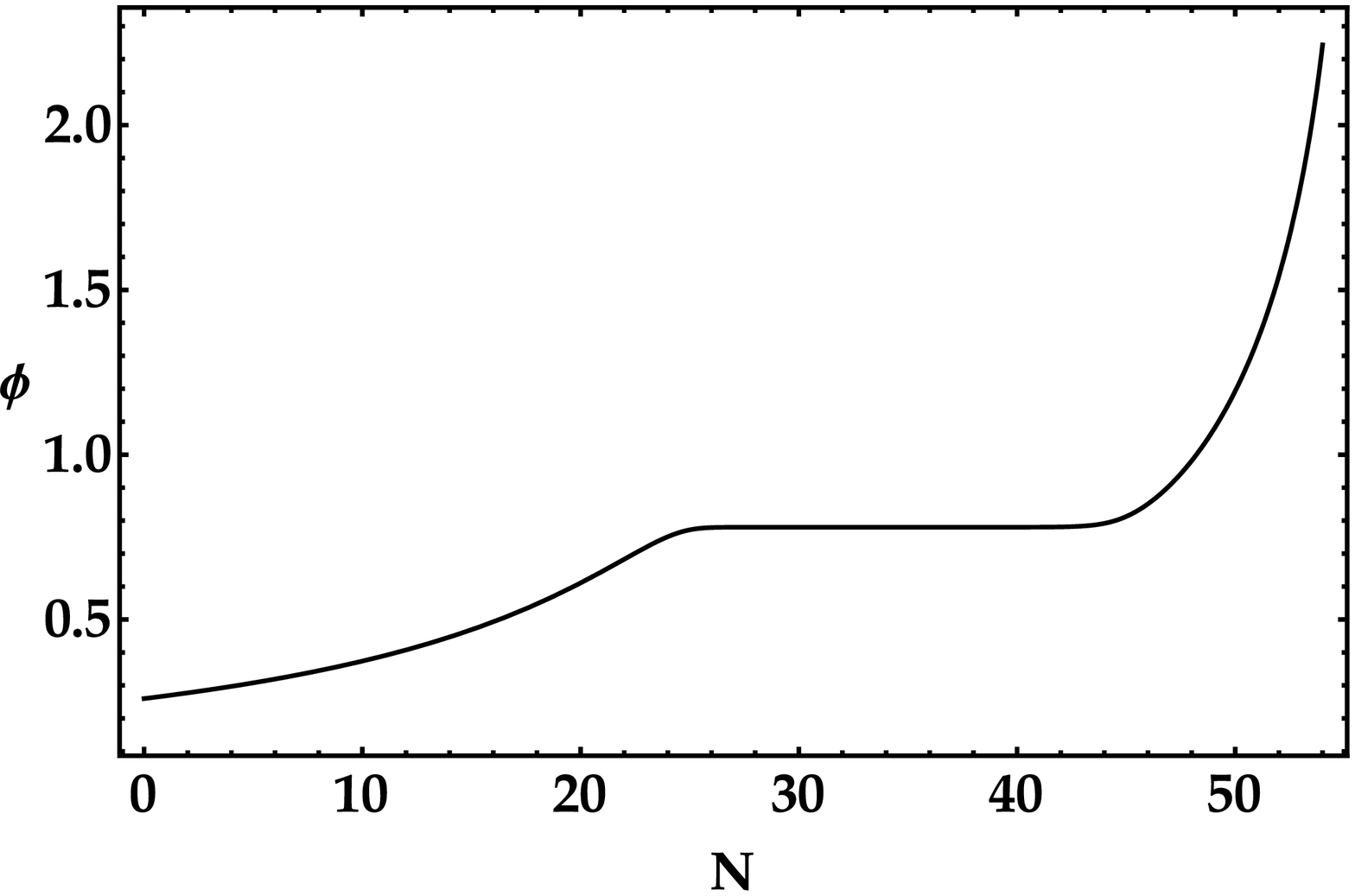}}\hspace{.1cm}
\subfigure[\label{fig-eps-n1}]{ \includegraphics[width=.45\textwidth]%
{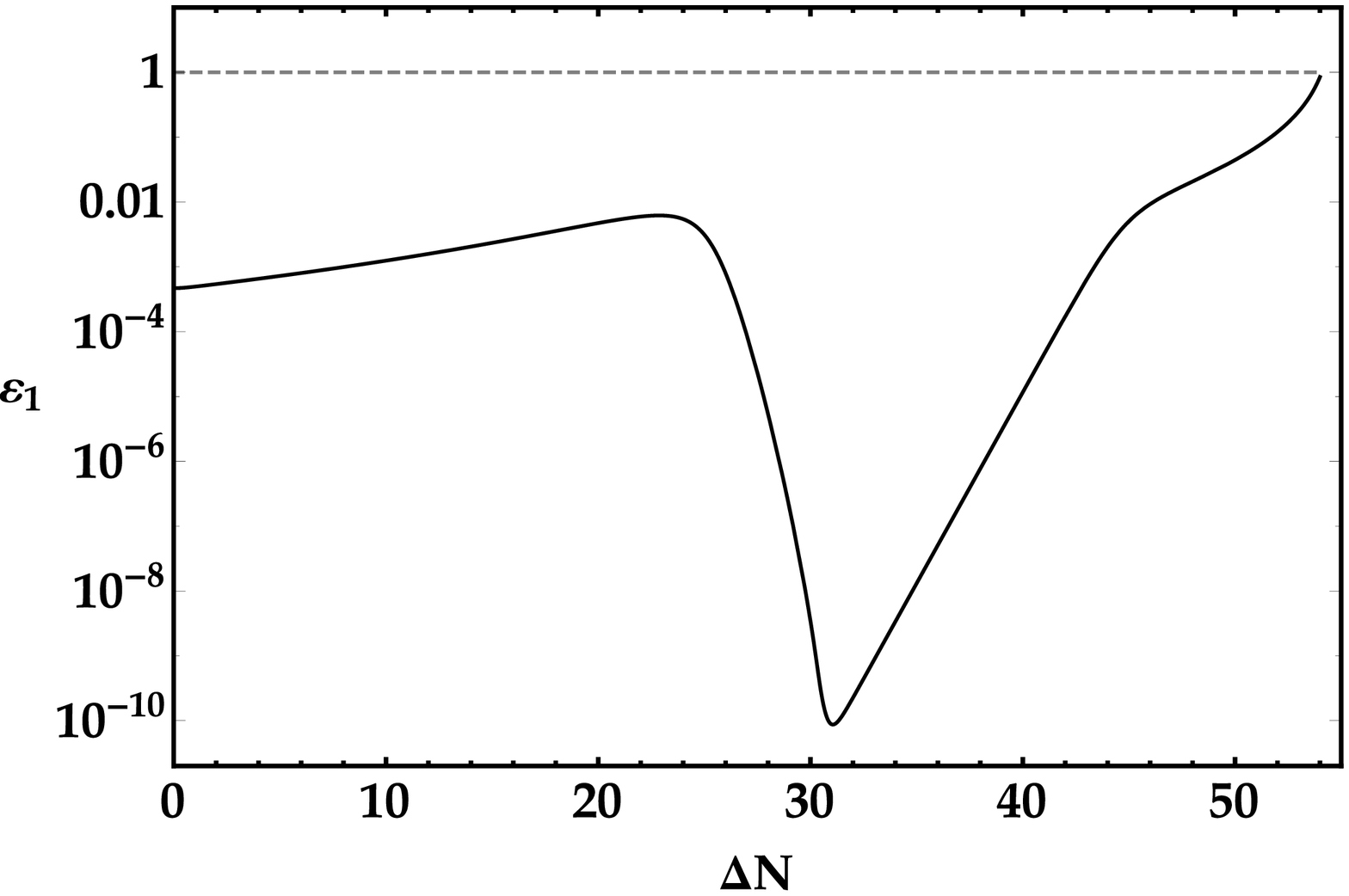}}
\centering
\subfigure[\label{fig-eta-n1}]{
 \includegraphics[width=.45\textwidth]%
{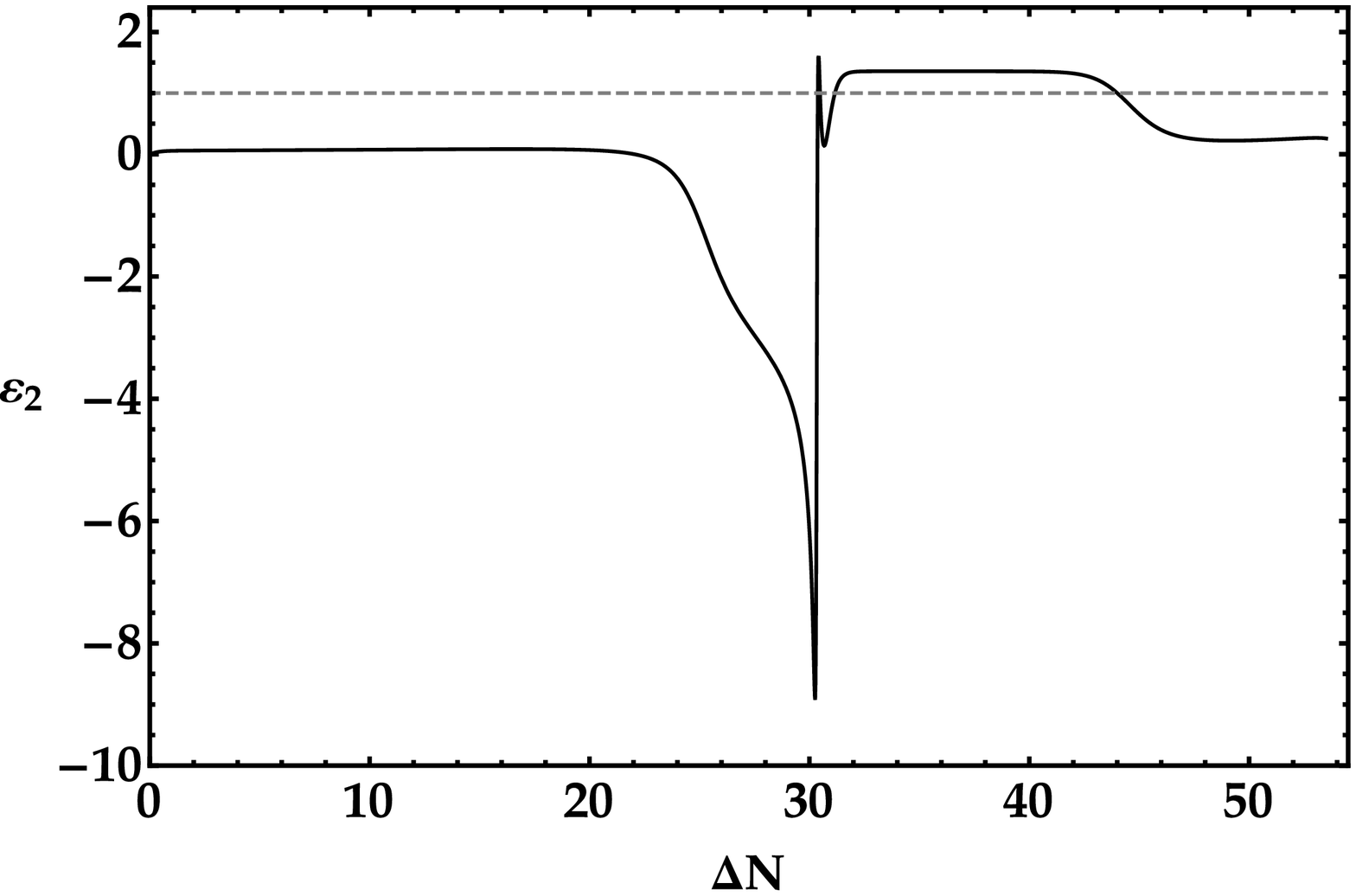}}
\end{minipage}
\caption{Same as Fig. \ref{linear}, but for the natural inflation model. The auxiliary parameters are given by the parameter set of case $A_{\rm N}$ in Table \ref{tab3} and we take $N_{*}=0$ at the horizon exit.
  }\label{linear}
\end{figure}

\begin{figure}[H]
\centering
\includegraphics[scale=0.5]{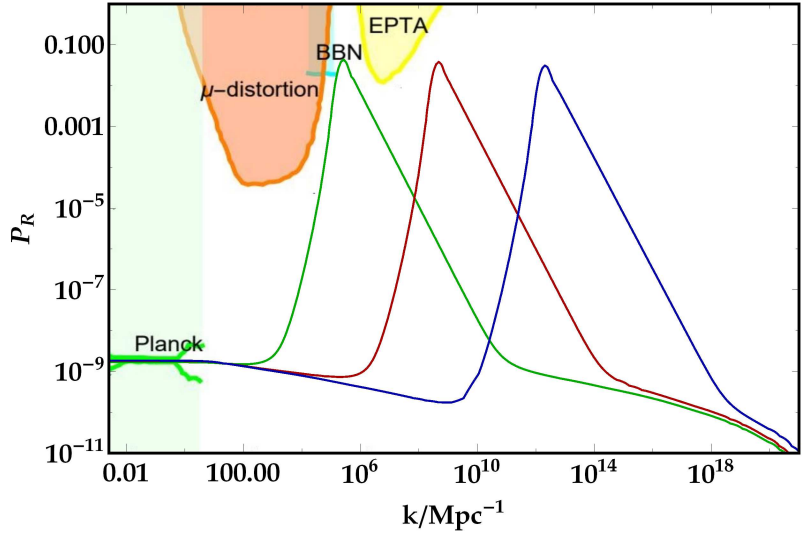}
\vspace{-0.5em}
\caption{Same as Fig. \ref{fig-pr-higgs}, but for the natural inflation model. The blue, red and green  lines are corresponding to the cases $A_{\rm N}$, $B_{\rm N}$ and $C_{\rm N}$, respectively. }
\label{fig-pr-nat}
\end{figure}

Table \ref{tab4} shows the values of power spectrum and the PBHs abundances corresponding to the three cases of Table \ref{tab3}. In Fig. \ref{fig-phi-n1}, evolution of the scalar field as a function of the $e$-fold number is plotted for the case $A_{\rm N}$ in Table \ref{tab3}. In this figure, the plateau-like region at $\phi=\phi_{c}$ leads to a severe enhancement of the primordial curvature perturbations (see Fig. \ref{fig-pr-nat}). In this region, the slow-roll condition is violated because  $|\varepsilon_2| > 1$, as shown in Fig. \ref{fig-eta-n1}.

Figure \ref{fig-pr-nat} shows the scalar power spectrum for the all cases of Table \ref{tab3}. For the PBHs formation, one needs the scalar power spectrum to increase seven orders at the small scales, and Fig. \ref{fig-pr-nat} exhibits such growth. Furthermore, Fig. \ref{fig-pr-nat} shows that the scalar power spectrum of this model is in good agreement with the observational bounds. These constraints include the observations of CMB $\mu$-distortion, big bang nucleosynthesis (BBN), and pulsar timing array (PTA) \cite{Inomata:2019-a,Inomata:2016,Fixsen:1996}.

\begin{table}[H]
  \centering
  \caption{The parameter sets for PBHs production in natural inflation model. The parameters $\phi_c$ and $c$ are in units of $M_{\rm{pl}}=1$. Here, $M=1.7$, $\alpha=-1$, and $f=1$.}
  \begin{tabular}{cccc}
  \hline
  Sets \quad &\quad $\phi_{c}$ \quad & \quad $d$\quad &$c$\quad \\ [0.5ex]
  \hline
  \hline
$\rm{Case~}A_{\rm N}$ \quad &\quad $0.74$ \quad &\quad $5.34\times 10^{9}$ \quad &\quad $1.9\times10^{-11}$ \quad\\[0.5ex]
  \hline
$\rm{Case~}B_{\rm N}$ \quad &\quad $0.41$ \quad &\quad $2.272\times 10^{9}$ \quad &\quad $1.3\times10^{-11}$ \quad\\[0.5ex]
  \hline
$\rm{Case~}C_{\rm N}$ \quad &\quad $0.28$ \quad &\quad $1.311\times 10^{9}$ \quad &\quad $1.07\times10^{-11}$ \quad\\[0.5ex]
  \hline
  \end{tabular}
  \label{tab3}
\end{table}

\begin{table}[H]
  \centering
  \caption{The values of $n_s$, $r$, $k_{\text{peak}}$, ${\cal P}_{\cal R}^\text{peak}$, $M_{\text{PBH}}^{\text{peak}}$ and $f_{\text{PBH}}^{\text{peak}}$ for the cases listed in Table \ref{tab3}.}
  \begin{tabular}{ccccccc}
  \hline
  Sets \quad & \quad $n_{s}$\quad &$r$\quad & \quad$k_{\text{peak}}/\text{\rm Mpc}^{-1}$ \quad & \quad ${\cal P}_{\cal R}^\text{peak}$\quad &\quad $M_{\text{PBH}}^{\text{peak}}/M_{\odot}$ \quad  & \quad $f_{\text{PBH}}^{\text{peak}}$\\ [0.5ex]
  \hline
  \hline
$\rm{Case~}A_{\rm N}$ \qquad &\quad $0.964$ \quad &\quad $0.007$ \quad &\quad $2.03\times10^{12}$ \quad &\quad $0.036$ \quad &\quad  $6.05\times 10^{-13}$ \quad &$0.92$ \\[0.5ex]
  \hline
$\rm{Case~}B_{\rm N}$ \quad &\quad $0.965$ \quad &\quad $0.006$ \quad &\quad $4.8\times10^{8}$ \quad &\quad $0.044$ \quad &\quad $1.60\times10^{-5}$ \quad & $0.035$ \\[0.5ex]
  \hline
$\rm{Case~}C_{\rm N}$ \quad &\quad $0.961$ \quad &\quad $0.005$ \quad &\quad $2.68\times10^{5}$ \quad &\quad $0.053$ \quad &\quad $34.8$ \quad &$0.0015$\\[0.5ex]
  \hline
  \end{tabular}
  \label{tab4}
\end{table}


\section{Reheating stage}\label{sec4}
After inflation, the scalar field starts to oscillate around the minimum value of the potential. Consequently, inflaton is decayed to the particles of the Standard Model. This process is known as the reheating, which joins the supercooled universe at the end of inflation to the thermalized universe at the radiation dominated (RD) epoch.

The decay rate of the inflaton can influence the reheating era duration. On the other hand, the reheating temperature $T_{\rm reh}$ can affect the decay rate of the inflaton. Hence, the large and small reheating temperatures lead to the short and prolonged reheating eras, respectively. The horizon re-entry can occur before the RD era if reheating phase has a prolonged period. Depending on the re-entry horizon era, the mathematical formalisms applied to determine the mass fraction and the energy density of GWs will be different. Here, we discuss whether PBHs produce in the RD era or reheating epoch \cite{mahbub:2020}.

In this paper, we use the method introduced in \cite{Dalianis:2019} to estimate the time of horizon re-entry. For a scale $k^{-1}$ that exits the horizon $\Delta N_{k}$ $e$-fold before the end of inflation, we can write
\begin{equation}\label{deltank}
\left( \frac{a_{k,\text{re}}}{a_{\text{end}}} \right)^{\frac{1}{2}(1+3w)}=e^{\Delta N_{k}},
\end{equation}
where $a_{k,\text{re}}$ denotes the scale factor at the horizon re-entry moment, and $w$  is the equation of state parameter. The number of $e$-folds between the end of inflation and horizon re-entry can be defined as
\begin{equation}\label{nk}
\tilde{N}_{k}\equiv \ln\left( \frac{a_{k,\text{re}}}{a_{\text{end}}} \right).
\end{equation}
Consequently, Eqs. (\ref{deltank}) and (\ref{nk}) can be related to each other as follows
\begin{equation}\label{deltaNtoN}
\tilde{N}_{k}=\left(\frac{2}{1+3w}\right)\Delta N_{k},
\end{equation}
where $w>-1/3$. The number of $e$-folds throughout the reheating period is
 $\tilde{N}_{\text{reh}}\equiv \ln\left( a_{\text{reh}}/a_{\text{end}} \right)=\left[3(1+w_{\text{reh}}) \right]^{-1} \ln\left(\rho_{\text{end}}/\rho_{\text{reh}} \right) $,
in which $a_{reh}$ shows the value of the scale factor at the end of reheating \cite{mahbub:2020,Dalianis:2019}. Also, $\rho_{\text{end}}=3H_{\text{end}}^{2}M_{p}^2$ is the energy density at the end of inflation epoch.

The values of $\tilde{N}_{k}$ and $\tilde{N}_{\text{reh}}$ can specify the time of horizon re-entry. If $\tilde{N}_{k}>\tilde{N}_{\text{reh}}$, the scale $k^{-1}$ re-enters during the RD era. On the other hand, the re-entering of the scale $k^{-1}$ with $\tilde{N}_{k}<\tilde{N}_{\text{reh}}$ occurs during the reheating stage. It is possible to assume $w_{\rm reh}\simeq 0$ because the reheating phase can be recognized as an early matter-dominated era. The amount of observable inflation $\Delta N$ is given by \cite{mahbub:2020}

\begin{equation}\label{deltaNfinal}
\Delta N\simeq 57.3+\frac{1}{4}\ln\left( \frac{\varepsilon_{*}V_{*}}{\rho_{\text{end}}} \right)-\frac{1}{4}\tilde{N}_{\text{reh}}~,
\end{equation}
where $\varepsilon_{*}$ and $V_{*}$ show the values of the first slow-roll parameter and the potential at the pivot scale, respectively.
For a scale such as $k^{-1}$, which re-enters the horizon just at the end of the reheating era, we have $\tilde{N}_{k}=\tilde{N}_{\text{reh}}$. Hence, $\Delta N_{\text{k}}$ and $\tilde{N}_{\text{reh}}$ can be associated with $\Delta N_{\text{k}}=\tilde{N}_{\text{reh}}/2$ in which we used Eq. (\ref{deltaNtoN}) and assumed $w_{\text{reh}}\simeq 0$. Therefore, the scale $k^{-1}$ re-enters horizon after reheating phase if $\Delta N_{\text{k}} > \tilde{N}_{\text{reh}}/2$.

Now for the both models studied in this paper, we can calculate the values of $\Delta N_{\text{k}}^{\rm peak}$ and $\tilde{N}_{\text{reh}}$, where $\Delta N_{\text{k}}^{\rm peak}$ describes the number of $e$-folds after the end of inflation until the scale $k^{-1}_{\text{peak}}$ re-enters the horizon. The co-moving wavenumber of the scale $k^{-1}$ that re-enters the horizon at the end of reheating phase, $k^{-1}_{\rm reh}$, is given by $k_{\rm reh}=e^{-\frac{\tilde{N}_{\rm reh}}{2}} k_{\rm end}$, where $k^{-1}_{\rm end}$ is the scale that exits the horizon at the end of inflation \cite{Dalianis:2019}.

\begin{figure}[H]
\begin{minipage}[b]{1\textwidth}
\subfigure[\label{fig-reh-higgs} ]{ \includegraphics[width=.46\textwidth]%
{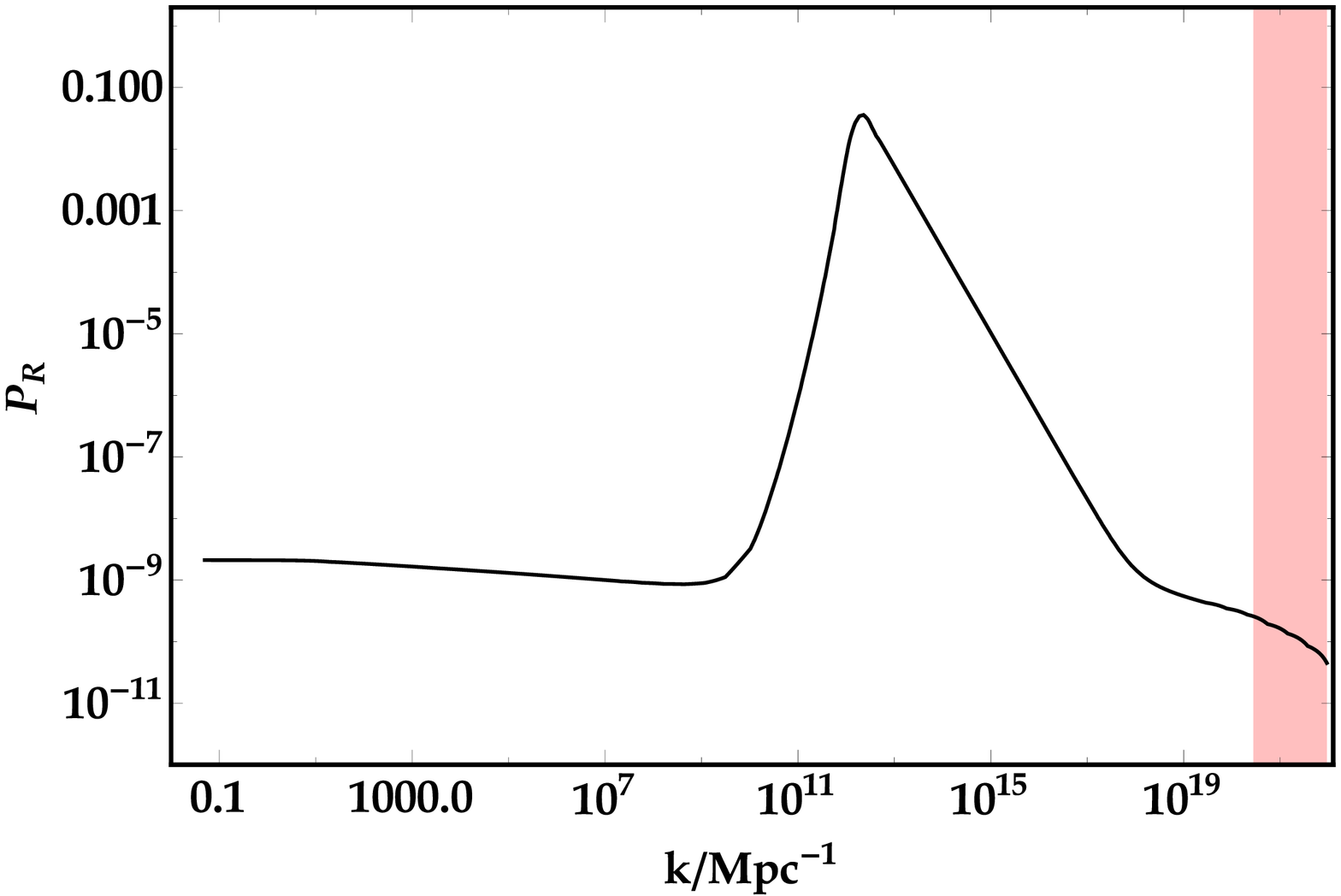}}\hspace{.1cm}
\subfigure[\label{fig-reh-nat}]{ \includegraphics[width=.46\textwidth]%
{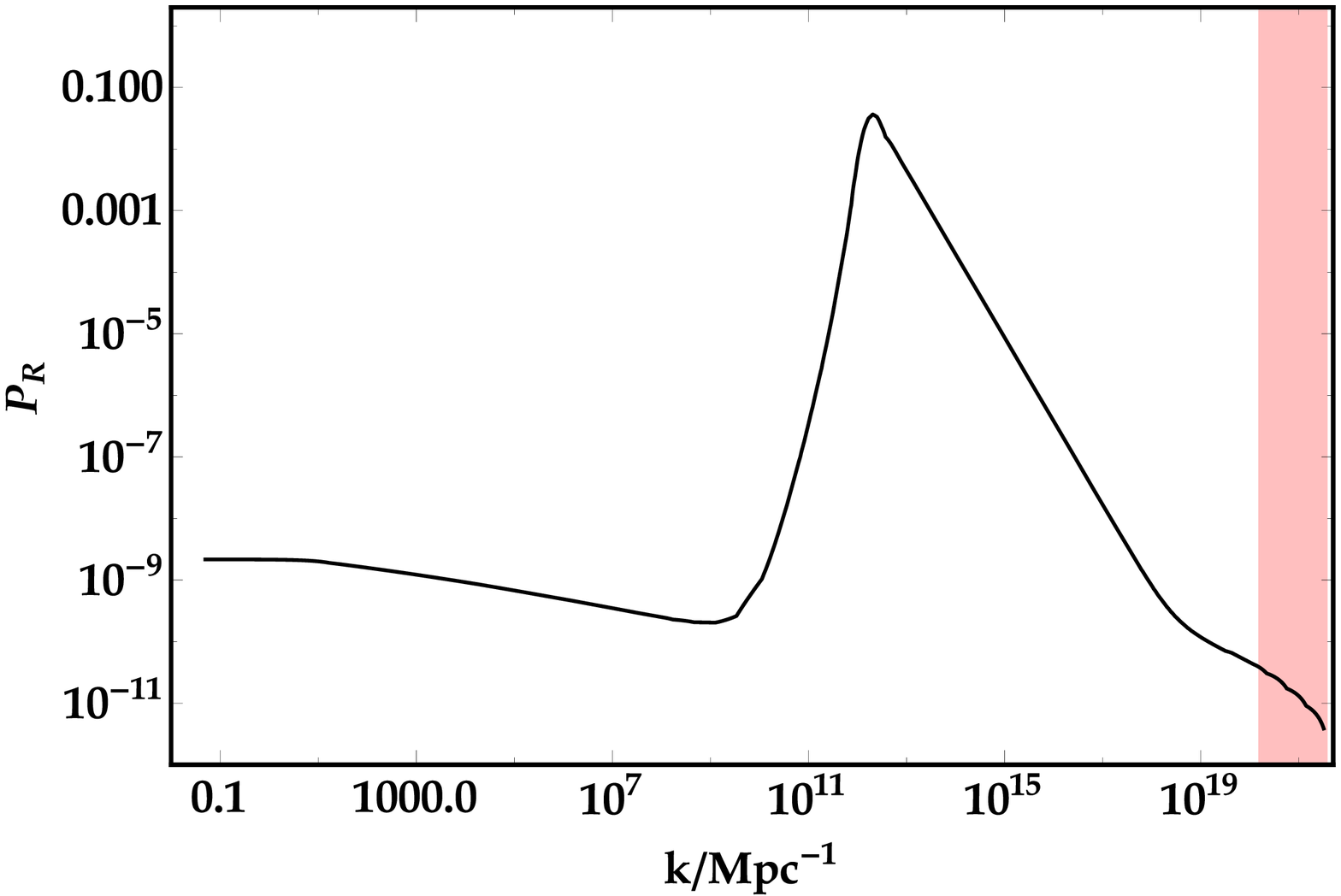}}
\end{minipage}
\caption{The scalar power spectrum for the cases (a) $A_{\rm H}$ and (b) $A_{\rm N}$, respectively, in quartic and natural inflation models. The shaded areas denote
the scales that re-enter the horizon during the reheating phase }\label{reh-fig}
\end{figure}

Our results in Table \ref{tab5} indicate that for the all cases in the both models we have $\Delta N_{\text{k}}^{\rm peak} > \tilde{N}_{\text{reh}}/2$. This means that in our models, the PBHs formation occurs in the RD era.
For instance, in Figs. \ref{fig-reh-higgs} and \ref{fig-reh-nat} we plot the scalar power spectrum associated with the cases $A_{\rm H}$ and $A_{\rm N}$, respectively.
The shaded regions describe the scales that re-enters the horizon during the reheating stage.
As shown in this figure, the duration of the reheating phase is short, and the peak scales re-enter the horizon after reheating epoch.
Hence, In the following section, we use the mathematical formalisms, which are verified in the RD era to calculate PBHs abundance and energy density of the induced GWs.
\begin{table}[H]
\centering
\caption{The values of $N_{\text{peak}}$, $\Delta N_{\text{k}}^{\rm peak}$ and $\tilde{N}_{\text{reh}}$ for the all cases in Tables \ref{tab1} and \ref{tab3}.}
\begin{tabular}{cccccc}
\hline
 Sets \quad & \quad  $\Delta N$\quad & $k_{\text{peak}}$                        \quad & \quad $N_{\text{peak}}$ \quad &\quad $\Delta N_{\text{k}}^{\rm peak}$ \quad & \quad$\tilde{N}_{\text{reh}}$ \quad \\ [0.5ex]
\hline
$\rm{Case~}A_{\rm H}$\quad & \quad 54.5 \quad & \quad $2.29\times 10^{12}$ \quad & \quad 31.6 \quad & \quad 16.42 \quad & \quad 6.98 \quad\\ \hline
$\rm{Case~}B_{\rm H}$ \quad & \quad 54.4 \quad & \quad $4.52\times 10^{8}$ \quad & \quad 23 \quad & \quad 22.16  \quad & \quad 7.44 \quad\\ \hline
$\rm{Case~}C_{\rm H}$ \quad & \quad 54.3 \quad & \quad $2.61\times 10^{5}$ \quad & \quad 15.5 \quad & \quad 26.56 \quad & \quad 7.85 \quad\\ \hline\hline
$\rm{Case~}A_{\rm N}$ \quad & \quad 54 \quad & \quad $2.03\times 10^{12}$ \quad & \quad 31.4 \quad & \quad 16.5  \quad & \quad 8.4 \quad\\ \hline
$\rm{Case~}B_{\rm N}$ \quad & \quad 53.5 \quad & \quad $4.8\times 10^{8}$ \quad & \quad 23 \quad & \quad 22.04  \quad & \quad 9.2  \quad\\ \hline
$\rm{Case~}C_{\rm N}$ \quad & \quad 53.5 \quad & \quad $2.68\times 10^{5}$ \quad & \quad 15.5 \quad & \quad 26.83  \quad & \quad 9.2  \quad\\ \hline
\end{tabular}
\label{tab5}
\end{table}

\section{Abundance of primordial black holes}\label{sec5}

During the RD era, when primordial curvature perturbations re-enter the horizon, the gravitational collapse may generate PBHs. The PBH mass at the production time is given by $M=\gamma M_{\rm H}$, where $M_{\rm H}$ is the horizon mass and $\gamma=0.2$ is the collapse efficiency parameter \cite{Inomata:2017,Sasaki:2018}. The current ratio of the PBH mass to the total DM reads \cite{Sasaki:2018}
\begin{equation}\label{fpbheq}
f_{\rm PBH}(M_{\rm PBH})=1.68\times 10^{8} \left(\frac{\gamma}{0.2} \right)^{\frac{1}{2}} \left(\frac{g_{*}}{106.75} \right)^{-\frac{1}{4}} \left(\frac{M_{\rm PBH}}{M_{\odot}} \right)^{-\frac{1}{2} }\beta(M_{\rm PBH}),
\end{equation}
where $g_{*}\simeq106.75$ is the effective degrees of freedom at the formation time of the PBH \cite{Motohashi:2017}. The mass fraction of PBH $\beta$ in Eq. (\ref{fpbheq}) can be estimated by \cite{Sasaki:2018,young:2014,harada:2013,Musco:2013,Germani:2019,Shibata:1999,Polnarev:2007,Musco:2009}
\begin{equation}\label{betaeq}
\beta(M_{\rm PBH})=\gamma \frac{\sigma_{M_{\rm PBH}}}{\sqrt{2\pi}\delta_{\rm th}}\exp{\left(-\frac{\delta_{\rm th}^{2}}{2 \sigma_{M_{\rm PBH}}^{2}} \right)},
\end{equation}
where $\delta_{\rm th}=0.4$ shows the threshold density contrast for PBH formation. Also, $\sigma_{M_{\rm PBH}}$ is the variance of density contrast at the comoving horizon scale and it is obtained as \cite{young:2014}
\begin{equation}\label{sigmaeq}
\sigma_{k}^{2}=\left(\frac{4}{9} \right)^{2} \int \frac{{\rm d}q}{q} W^{2}(q/k)(q/k)^{4} P_{\cal R}(q),
\end{equation}
where
$W(x)=\exp{\left(-x^{2}/2 \right)} $
is the Gaussian window function.
It is possible to associate the mass of the PBHs and the corresponding wavenumber as \cite{Motohashi:2017,mishra:2020,Sasaki:2018}
\begin{equation}\label{masseq}
M_{\rm PBH}=1.13\times 10^{15}\left(\frac{\gamma}{0.2} \right) \left(\frac{g_{*}}{106.75} \right)^{-\frac{1}{6}}\left(\frac{k_{\rm PBH}}{k_{*}} \right)^{-2} M_{\odot}.
\end{equation}

\begin{figure}[H]
\begin{minipage}[b]{1\textwidth}
\subfigure[\label{fig-fpbh-higgs} ]{ \includegraphics[width=.46\textwidth]%
{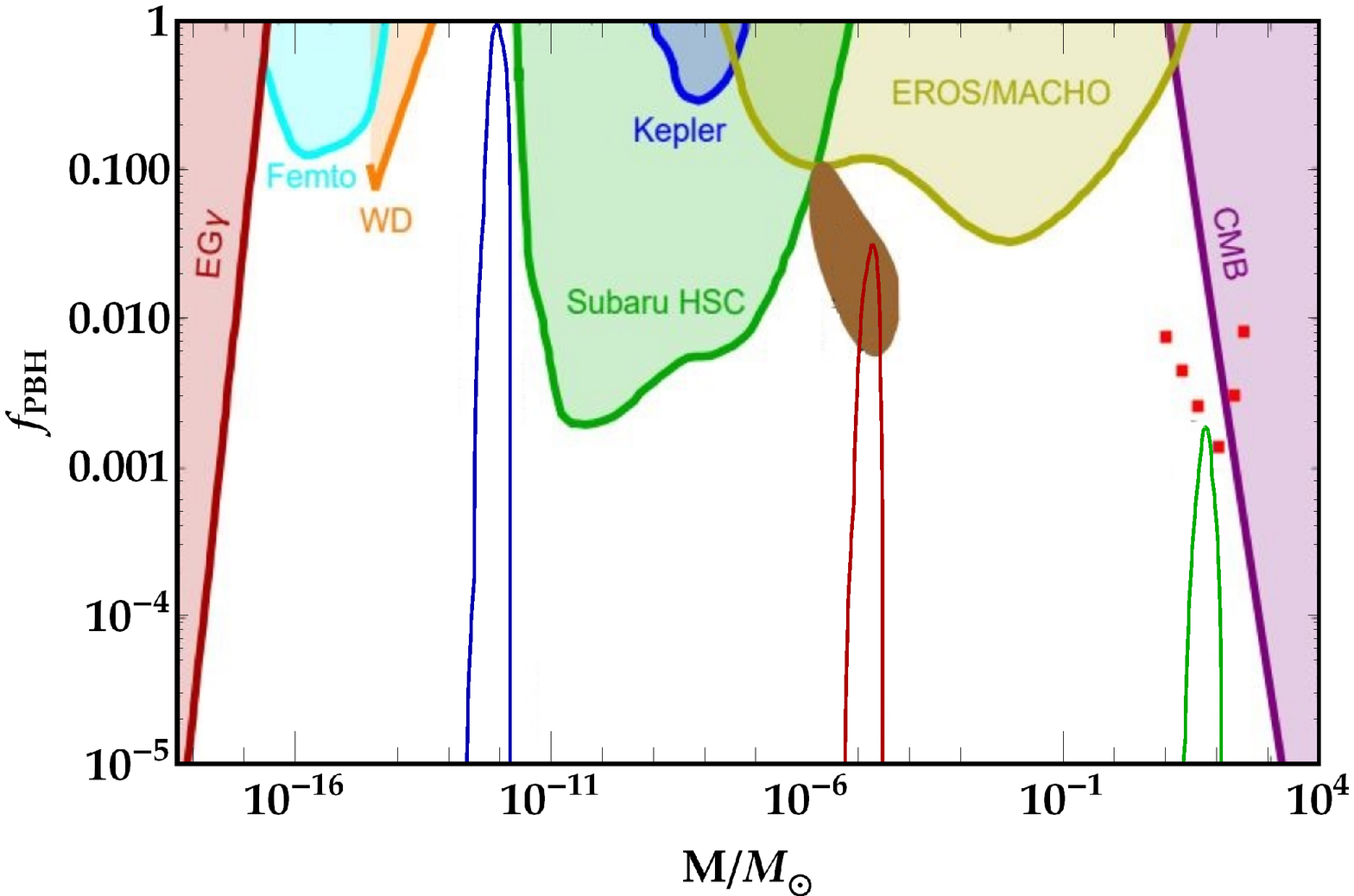}}\hspace{.1cm}
\subfigure[\label{fig-fpbh-nat}]{ \includegraphics[width=.46\textwidth]%
{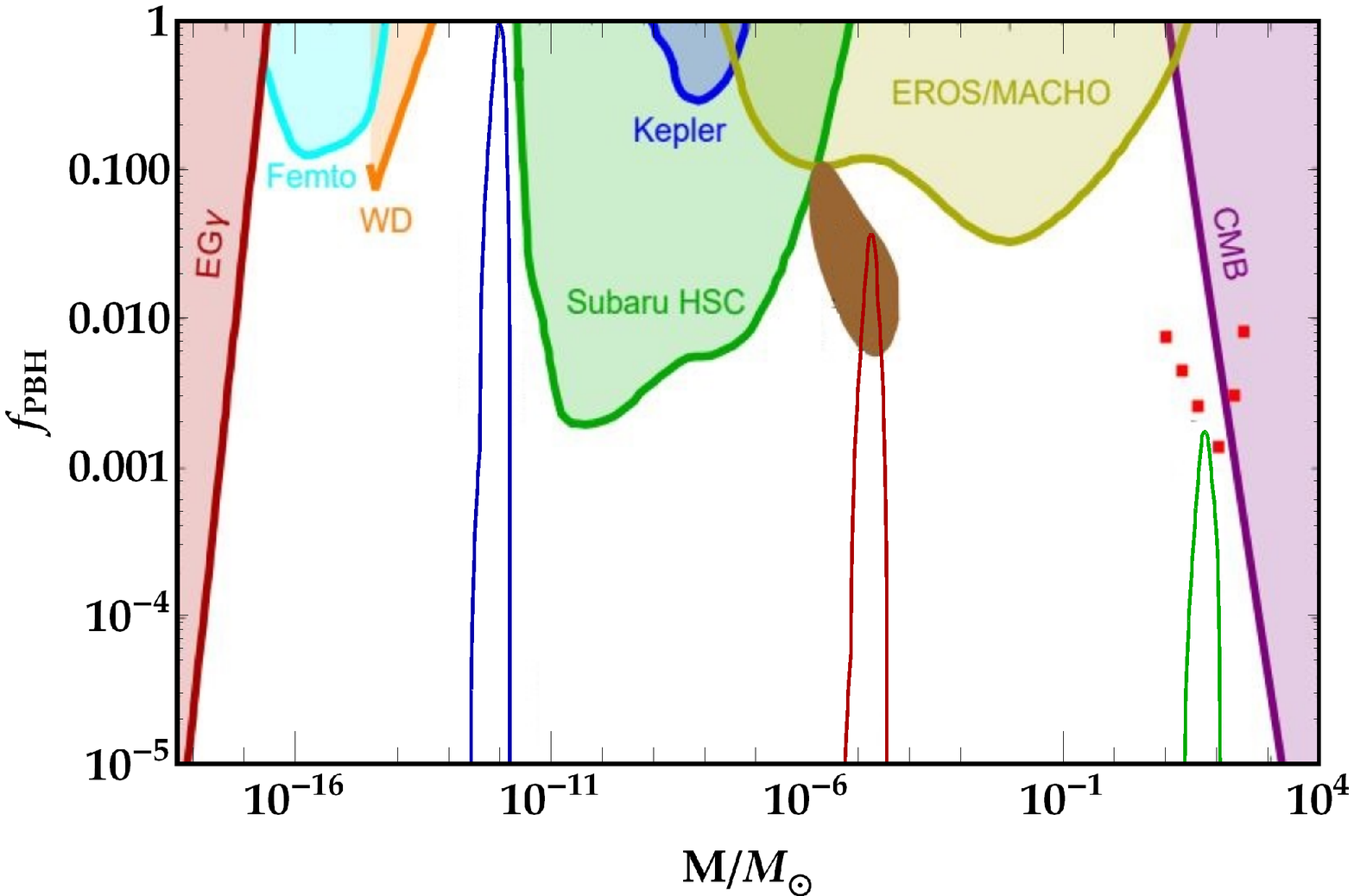}}
\end{minipage}
\caption{(a) The PBH abundance $f_{\rm PBH}$ versus the mass $M$ in the quartic inflation model for the cases $A_{\rm H}$ (blue line), $B_{\rm H}$ (red line), and $C_{\rm H}$ (green line). (b) Same as Fig. \ref{fig-fpbh-higgs} but in the natural inflation model for the cases $A_{\rm N}$ (blue line), $B_{\rm N}$ (red line), and $C_{\rm N}$ (green line). The observational constraints on PBH abundance are shown by the shaded areas. The brown shaded area illustrates the allowed region of PBH abundance in the OGLE data \cite{OGLE}. The red dots show the upper bound on the PBH abundance due to the upper limit on the LIGO event merger rate \cite{Ali:2017}. Other shaded regions present the recent observational constraints containing accretion constraints from CMB \cite{CMB-a,CMB-b}, with EROS/MACHO (EROS/MACHO) \cite{EROS}, with the Kepler satellite (Kepler) \cite{kepler}, microlensing events with Subaru HSC (Subaru HSC) \cite{HSC}, white dwarf explosion (WD) \cite{WD}, femtolensing of gamma-ray burst (Femto) \cite{femto} and extragalactic gamma rays from PBH evaporation (EG$\gamma$) \cite{EGG}.}\label{fpbh-figs}
\end{figure}

With the help of Eqs. (\ref{fpbheq}) and (\ref{masseq}), one can estimate the abundance and the mass of PBHs. For the both quartic and natural inflationary models, the results of $M_{\text{PBH}}^{\text{peak}}$ and $f_{\text{PBH}}^{\text{peak}}$ calculated at the peak scale of the PBH formation are listed in Tables \ref{tab2} and \ref{tab4}. Also in Fig. \ref{fpbh-figs}, we plot variations of $f_{\text{PBH}}$ versus $M_{\text{PBH}}$ for the both models. In this figure, the shaded areas reveal the observational bounds on the abundance of PBH. Figure \ref{fpbh-figs} shows that (i) the PBHs with masses $\sim10^{-13}M_\odot$, $\sim10^{-5}M_\odot$, and $\sim10M_\odot$ which are compatible with the observations can be produced in the aforementioned models. (ii) Surprisingly enough is that for the cases $A_{\rm H}$ and $A_{\rm N}$, we obtain $f_{\rm PBH}^{\rm peak}=0.91$ and $f_{\rm PBH}^{\rm peak}=0.92$, respectively, which can explain most of DM in the universe. (iii) For the cases  ${\rm B_{H}}$  and ${\rm B_{N}}$, the abundance peak of the PBHs can grow up to $f_{\text{PBH}}^{\text{peak}}= 0.031$ and $f_{\text{PBH}}^{\text{peak}}= 0.035$, respectively, for the mass scales $1.82\times10^{-5}M_\odot$ and $1.60\times10^{-5}M_\odot$. The results obtained for the cases ${\rm B_{H}}$  and ${\rm B_{N}}$ are located at the allowed region of microlensing events in the OGLE data.
(iv) For the cases ${\rm C_{H}}$ and ${\rm C_{N}}$, the peaks of the PBH abundance are located at $36.9M_{\odot}$ and $34.8M_{\odot}$ and enhance to $f_{\text{PBH}}^{\text{peak}}= 0.0017$ and $f_{\text{PBH}}^{\text{peak}}= 0.0015$, respectively. This result is compatible with the upper limit bounds on the LIGO merger rate and consequently can describe BHs observed by LIGO and Virgo collaboration.

\section{Induced Gravitational Waves}\label{sec6}

\begin{figure}[H]
\begin{minipage}[b]{1\textwidth}
\subfigure[\label{fig-omega-higgs} ]{ \includegraphics[width=.49\textwidth]%
{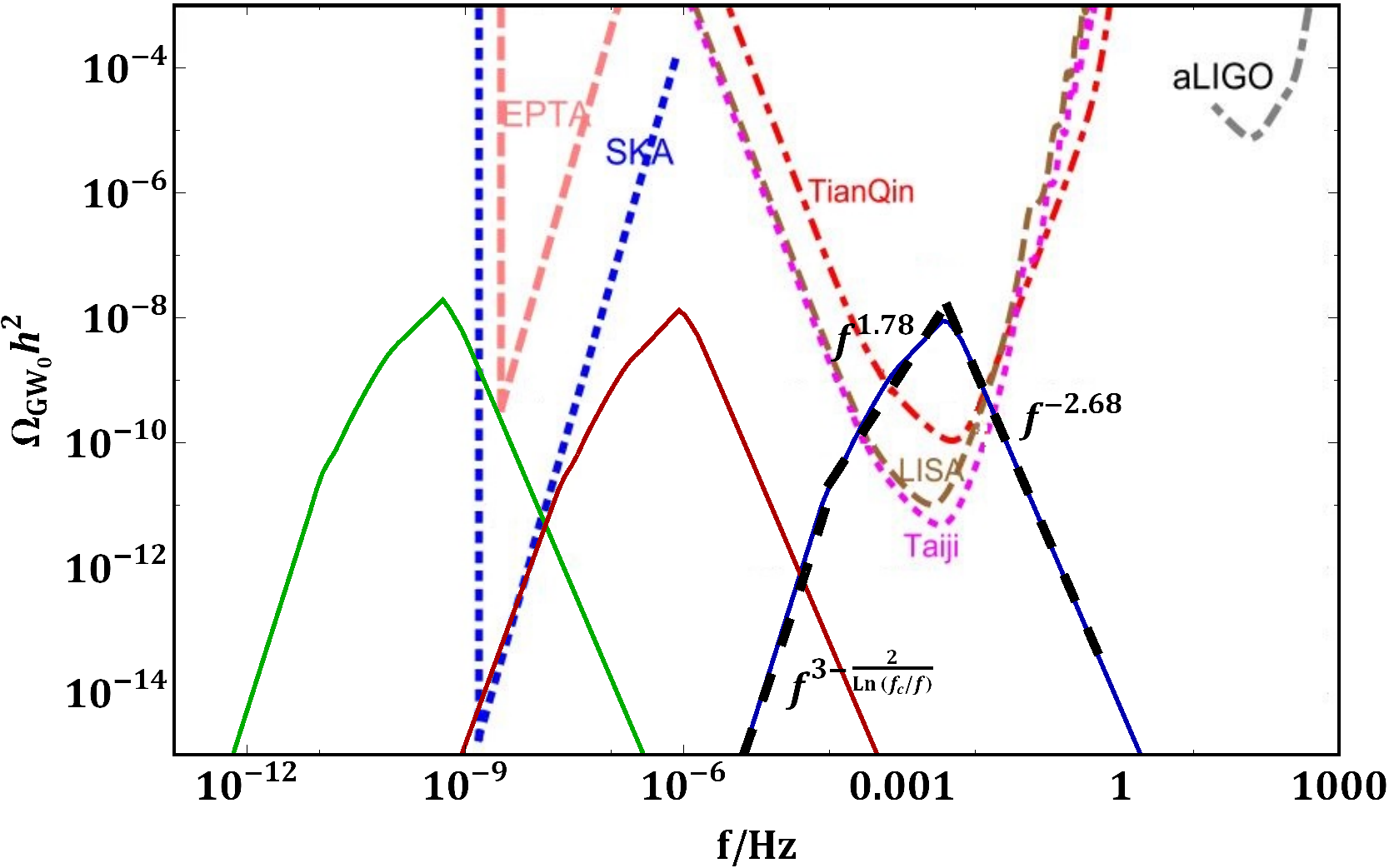}}\hspace{0.01cm}
\subfigure[\label{fig-omega-nat}]{ \includegraphics[width=.49\textwidth]%
{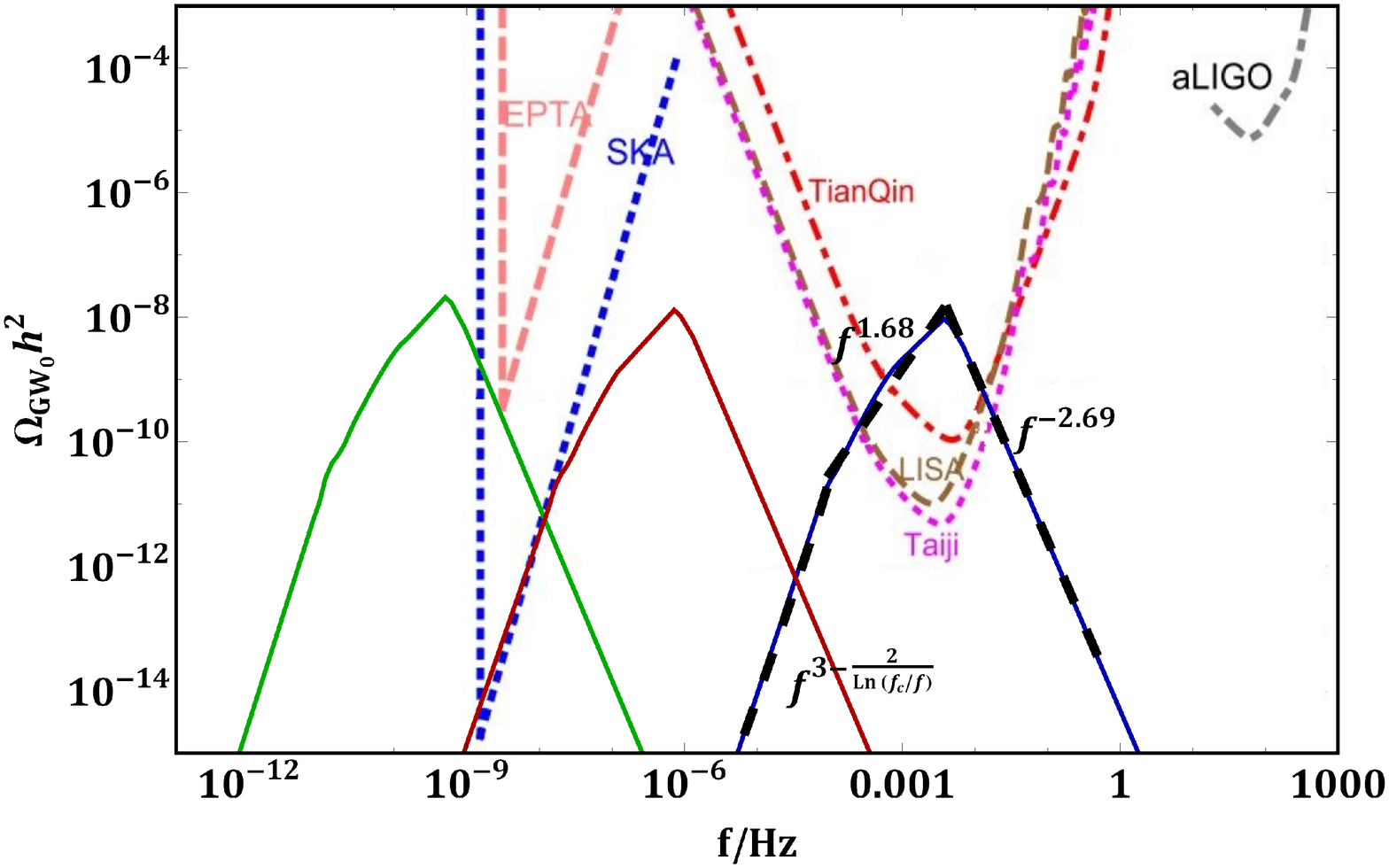}}
\end{minipage}
\caption{(a) The induced GWs energy density parameter $\Omega_{\rm GW{,0}}$ versus the frequency in the quartic inflation model for the cases $A_{\rm H}$ (blue line), $B_{\rm H}$ (red line), and $C_{\rm H}$ (green line). (b) Same as Fig. \ref{fig-omega-higgs} but in the natural inflation model  for the cases $A_{\rm N}$ (blue line), $B_{\rm N}$ (red line), and $C_{\rm N}$ (green line). The  black dashed line denotes the broken power-law behaviour of $\Omega_{GW}$. The dashed curves describe the sensitivity of GWs observatories, such as the European PTA (EPTA) \cite{EPTA-a,EPTA-b,EPTA-c,EPTA-d}, the Square Kilometer Array (SKA) \cite{ska}, the Advanced Laser Interferometer Gravitational Wave Observatory (aLIGO) \cite{ligo-a,ligo-b}, the Laser Interferometer Space Antenna (LISA) \cite{lisa,lisa-a}, Taiji \cite{taiji} and TianQin \cite{tianqin}.}\label{omega-figs}
\end{figure}

The induced GWs can produce together with PBHs at the horizon re-entry of the primordial density perturbations  \cite{Matarrese:1998,Mollerach:2004,Saito:2009,Garcia:2017,Cai:2019-a,Cai:2019-b,Cai:2019-c,Bartolo:2019-a,Bartolo:2019-b,Wang:2019,Fumagalli:2020b,Domenech:2020a,Domenech:2020b,Hajkarim:2019,Kohri:2018,Xu:2020}.
The induced GWs can be tested by the GWs observatory like LISA \cite{lisa}. To investigate the induced GWs, we start with the perturbed FRW metric in the conformal Newtonian gauge which is written as \cite{Ananda:2007}
\begin{eqnarray}
ds^2=a(\eta)^2\left[ -(1+2\Psi)d\eta^2 +\left((1-2\Psi)\delta_{ij}+\frac{h_{ij}}{2} \right)dx^idx^j \right],
\end{eqnarray}
where $a$ is the scale factor and $\eta$ represents the conformal time. Also $\Psi$ denotes the first-order scalar perturbations, and
$h_{ij}$ is the perturbation of the second-order transverse-traceless tensor. After inflation, the universe should be thermalized in the reheating era. Hence, the inflaton will be decayed into the light particles, and the universe will be dominated by radiation, consequently. In the RD era, the effect of the inflation field has almost ignorable on the cosmic evolution. Accordingly, to study the scalar induced GWs during the RD era, we can use the standard Einstein equation. So, the equation of motion for second-order tensor perturbations $h_{ij}$ satisfies \cite{Ananda:2007,Baumann:2007}
\begin{eqnarray}\label{EOM_GW}
h_{ij}^{\prime\prime}+2\mathcal{H}h_{ij}^\prime - \nabla^2 h_{ij}=-4\mathcal{T}^{lm}_{ij}S_{lm}\;,
\end{eqnarray}
where  $\mathcal{H}\equiv a^{\prime}/a$ is the conformal Hubble parameter. The quantity $\mathcal{T}^{lm}_{ij}$ denotes the transverse-traceless projection operator, and $S_{ij}$, which is the GW source term, is obtained as
\begin{eqnarray}
S_{ij}=4\Psi\partial_i\partial_j\Psi+2\partial_i\Psi\partial_j\Psi-\frac{1}{\mathcal{H}^2}\partial_i(\mathcal{H}\Psi+\Psi^\prime)\partial_j(\mathcal{H}\Psi+\Psi^\prime)\; .
\end{eqnarray}
During the RD era, the scalar metric perturbation $\Psi$ takes the form \cite{Baumann:2007}
\begin{eqnarray}
\Psi_k(\eta)=\psi_k\frac{9}{(k\eta)^2}\left(\frac{\sin(k\eta/\sqrt{3})}{k\eta/\sqrt{3}}-\cos(k\eta/\sqrt{3}) \right)\;,
\end{eqnarray}
where $k$ indicates the comoving wavenumber, and the primordial perturbation $\psi_k$ is given by
\begin{eqnarray}
\langle \psi_{\bf k}\psi_{ \tilde{\bf k}}  \rangle = \frac{2\pi^2}{k^3}\left(\frac{4}{9}\mathcal{P}_\mathcal{R}(k)\right)\delta(\bf{k}+ \tilde{\bf k})\;.
\end{eqnarray}
The energy density of the induced GWs in the RD era can be estimated by \cite{Kohri:2018}
\begin{eqnarray}\label{OGW}
&\Omega_{\rm{GW}}(\eta_c,k) = \frac{1}{12} {\displaystyle \int^\infty_0 dv \int^{|1+v|}_{|1-v|}du } \left( \frac{4v^2-(1+v^2-u^2)^2}{4uv}\right)^2\mathcal{P}_\mathcal{R}(ku)\mathcal{P}_\mathcal{R}(kv)\left( \frac{3}{4u^3v^3}\right)^2 (u^2+v^2-3)^2\nonumber\\
&\times \left\{\left[-4uv+(u^2+v^2-3) \ln\left| \frac{3-(u+v)^2}{3-(u-v)^2}\right| \right]^2  + \pi^2(u^2+v^2-3)^2\Theta(v+u-\sqrt{3})\right\}\;,
\end{eqnarray}
where $\Theta$ is the Heaviside theta function, and $\eta_{c}$ is the time in which the growing of $\Omega_{\rm{GW}}$ is stopped. Also the scalar power spectrum $\mathcal{P}_\mathcal{R}$ is estimated by the MS equation (\ref{MS_Eq}). The current energy density of the induced GWs can be obtained as follows \cite{Inomata:2019-a}
\begin{eqnarray}\label{OGW0}
\Omega_{\rm{GW},0}h^2 = 0.83\left( \frac{g_c}{10.75} \right)^{-1/3}\Omega_{\rm{r},0}h^2\Omega_{\rm{GW}}(\eta_c,k)\;,
\end{eqnarray}
where $\Omega_{\rm{r},0}h^2\simeq 4.2\times 10^{-5}$ is the radiation density parameter at the present time. Also $g_c\simeq106.75$ indicates the effective degrees of freedom in the energy density at $\eta_c$. The frequency and comoving wavenumber can be related as follows
\begin{eqnarray}\label{k_to_f}
f=1.546 \times 10^{-15} \left(\frac{k}{{\rm Mpc}^{-1}}\right){\rm Hz}.
\end{eqnarray}
Now, with numerical solving of Eq. (\ref{OGW}) and using (\ref{OGW0})-(\ref{k_to_f}) one can estimate the present energy density of induced GWs. The results of $\Omega_{\rm{GW},0}$ for the quartic and natural models are shown in Fig. \ref{omega-figs}. The blue, red, and green curves correspond to the PBHs with the mass in the order of $ 10^{-13}M_\odot$, $ 10^{-5}M_\odot$, and $ 10M_\odot$, respectively. The shape and the amplitude of the  current energy spectra of the induced GWs are almost the same for all cases, but the peak frequencies are different as depicted in Figs. \ref{fig-omega-higgs} and \ref{fig-omega-nat}.

Figure \ref{omega-figs} shows that (i) for the cases $A_{\rm H}$ and $A_{\rm N}$ with $M \sim {\cal O}(10^{-13})M_\odot$, the peak frequency of $\Omega_{GW,0}$ is around the $\rm{mHz}$ band, which means both of them can place in the sensitive region of space-based observatories like LISA, $\rm{TianQin}$, and Taiji.
(ii) In the cases $B_{\rm H}$ and $B_{\rm N}$ with PBH mass around $10^{-5}M_\odot$, the $\Omega_{GW,0}$ has a peak in $f \sim 10^{-6} \rm{Hz}$ band. (iii) The energy density of induced GWs produced from the cases $C_{\rm H}$ and $C_{\rm N}$ has a peak in frequencies in the order of $10^{-10} \rm{Hz}$. Hence, the cases $B_{\rm H}$, $C_{\rm H}$, $B_{\rm N}$, and $C_{\rm N}$ can be tested by SKA observation.

Recent studies confirm that the energy density can be parameterized as  $\Omega_{\rm GW} (f) \sim f^{n} $ \cite{Xu:2020,Fu:2020,Kuro:2018}.
In this regard, for the case $A_{\rm H}$ we estimate $\Omega_{\rm GW} \sim f^{1.78}$ for $f<f_{c}=3.54\times 10^{-3}{\rm Hz}$, and $\Omega_{\rm GW} \sim f^{-2.69}$ for $f>f_{c}$. Also, for the case $A_{\rm N}$, our calculations indicate that $\Omega_{\rm GW} \sim f^{1.68}$ for $f<f_{c}=3.73\times 10^{-3}{\rm Hz}$, and $\Omega_{\rm GW} \sim f^{-2.69}$ for $f>f_{c}$. In addition, the results in both cases confirm $\Omega_{\rm GW} \sim f^{3-2/\ln(f_c/f)}$ in the infrared limit $f\ll f_{c}$, which is entirely compatible with the results of \cite{Yuan:2020,shipi:2020}.

\section{Conclusions}\label{sec7}
In this paper, we investigated the possibility of PBHs formation in the context of inflation with field-dependent kinetic term for two types of the inflationary potentials (i.e. quartic and natural potentials). Utilizing a proper kinetic term may increase the primordial perturbations to ${\cal O}(10^{-2})$ at small scales, which is required for PBHs formation. On the other hand, the value of the $P_{\cal R}(k)$ should be consistent with the Planck observation at the pivot scale. In our model, we use a kinetic function that contains two parts as $G(\phi)=g_I(\phi)\big(1+g_{II}(\phi)\big)$. The first part $g_{I}(\phi)$ guarantees that the model is compatible with the Planck measurements at the CMB scale. Moreover, the fine-tuning of the parameters of the second section $g_{II}(\phi)$ can severely increase the $P_{\cal R}(k)$ at small scales.

We find three sets of parameters for each potential which are listed in Tables \ref{tab1} and \ref{tab3}. Our calculations demonstrate that the scales re-enter the horizon after the reheating stage. Therefore, we can apply the mathematical formalism that is valid in the RD epoch.
In the following, we estimated the PBHs abundance for all cases in Tables \ref{tab2} and \ref{tab4}. In our model, the noteworthy sets are $A_{\rm H}$ and $A_{\rm N}$, which their $f_{\rm PBH}$ are in order $ {\cal O}(1)$ and can explain the total DM in the universe. For the cases $B_{\rm H}$ and $B_{\rm N}$, the peak values of $f_{\rm PBH}$ are as $ {\cal O}(10^{-2})$, which locate in the allowed region of ultrashort timescale in OGLE data. Also, the cases of $C_{\rm H}$ and $C_{\rm N}$ are compatible with the upper limit of LIGO, and  $f_{\rm PBH}$ grows to $ {\cal O}(10^{-3})$.

In the last section, we studied the propagation of secondary GWs in our model. For the cases $A_{\rm H}$ and $A_{\rm N}$, the peaks of $\Omega_{\rm GW}$ are formed in the mHz frequency range and can be examined by the observations of LISA, Taiji, and TianQin. Moreover, as shown in the Fig. \ref{omega-figs}, in the cases $B_{\rm H}$, $C_{\rm H}$, $B_{\rm N}$, and $C_{\rm N}$ the GWs energy density parameter can be examined by the SKA observation. In addition, our numerical calculations expose that the GWs energy density parameter can be parameterized as a power-law function  $\Omega_{\rm GW} (f) \sim f^{n} $. The results show that $\Omega_{\rm GW} \sim f^{1.78}$ for $f<f_{c}=3.54\times 10^{-3}{\rm Hz}$, and $\Omega_{\rm GW} \sim f^{-2.69}$ for $f>f_{c}$ in the case of $A_{\rm H}$, and  $\Omega_{\rm GW} \sim f^{1.68}$ for $f<f_{c}=3.73\times 10^{-3}{\rm Hz}$, and $\Omega_{\rm GW} \sim f^{-2.69}$ for $f>f_{c}$ in the case of $A_{\rm N}$. Also, in the infrared limit $f\ll f_{c}$, for the both cases, the power index of the GWs energy density satisfies the relation $n=3-2/\ln(f_c/f) $, which is well consistent with the analytical result obtained in \cite{Yuan:2020,shipi:2020}.

\subsection*{Acknowledgements}
The authors thank the referee for his/her valuable comments.



\begin{thebibliography}{0}
\expandafter\ifx\csname natexlab\endcsname\relax\def\natexlab#1{#1}\fi
\expandafter\ifx\csname bibnamefont\endcsname\relax
  \def\bibnamefont#1{#1}\fi
\expandafter\ifx\csname bibfnamefont\endcsname\relax
  \def\bibfnamefont#1{#1}\fi
\expandafter\ifx\csname citenamefont\endcsname\relax
  \def\citenamefont#1{#1}\fi
\expandafter\ifx\csname url\endcsname\relax
  \def\url#1{\texttt{#1}}\fi
\expandafter\ifx\csname urlprefix\endcsname\relax\def\urlprefix{URL }\fi
\providecommand{\bibinfo}[2]{#2}
\providecommand{\eprint}[2][]{\url{#2}}

\end{thebibliography}


\begin{thebibliography}{100}
\bibitem{Hawking:1971}
S. Hawking, Mon. Not. R. Astron. Soc. {\bf 152}, 75 (1971).

\bibitem{Carr:1974}
B. J. Carr and S.W. Hawking, Mon. Not. R. Astron. Soc. {\bf 168}, 399 (1974).

\bibitem{Abbott:2016-a}
B. P. Abbott et al. (LIGO Scientific Collaboration and Virgo Collaboration), Phys. Rev. Lett.
{\bf 116}, 061102 (2016).
%
\bibitem{Abbott:2016-b}
B. P. Abbott et al. (LIGO Scientific Collaboration and Virgo Collaboration), Phys. Rev. Lett.
{\bf 116}, 241103 (2016).
%
\bibitem{Abbott:2017-a}
 B. P. Abbott et al. (LIGO Scientific Collaboration and Virgo Collaboration), Phys. Rev. Lett.
{\bf 116}, 221101 (2017).
%
\bibitem{Abbott:2017-b}
B. P. Abbott et al. (LIGO Scientific Collaboration and Virgo Collaboration), Astrophys. J.
{\bf 851}, L35 (2017).
%
\bibitem{Abbott:2017-c}
B. P. Abbott et al. (LIGO Scientific Collaboration and Virgo Collaboration), Phys. Rev. Lett.
{\bf 119}, 141101 (2017).
%
%
\bibitem{Ivanov:1994}
P. Ivanov, P. Nasselsky and I.D. Novikov, Phys. Rev. D {\bf 50}, 7173 (1994)
%
\bibitem{Khlopov1:2005}
 M. Yu. Khlopov, S. G. Rubin, and A. S. Sakharov, Astropart. Phys.  {\bf 23}, 265 (2005).
%
\bibitem{Frampton:2010}
P. H. Frampton, M. Kawasaki, F. Takahashi, and T. T. Yanagida, JCAP {\bf 1004}, 023 (2010).
%
\bibitem{Belotsky:2014}
K. M. Belotsky et al. Mod. Phys. Lett. A {\bf 29}, 1440005 (2014).
%
\bibitem{Clesse:2015}
 S. Clesse and J. GarcBellido, Phys. Rev. D {\bf 92}, 023524 (2015).
%
\bibitem{Carr:2016}
B. Carr, F. Kuhnel, and M. Sandstad, Phys. Rev. D {\bf 94}, 083504 (2016).
%
\bibitem{Inomata:2017}
K. Inomata, M. Kawasaki, K. Mukaida, Y. Tada, and T. T. Yanagida, Phys. Rev. D {\bf 96}, 043504 (2017).
%
\bibitem{Sasaki:2016}
M. Sasaki, T. Suyama, T. Tanaka, and S. Yokoyama, Phys. Rev. Lett. {\bf 117}, 061101 (2016).
%


\bibitem{OGLE}
H. Niikura, M. Takada, S. Yokoyama, T. Sumi, and S. Masaki, Phys. Rev. D \textbf{99}, 083503 (2019).
%
\bibitem{Ali:2017}
Y. Ali-Ha{\"i}moud, E. D. Kovetz, and M. Kamionkowski, Phys. Rev. D \textbf{96}, 123523 (2017).
%
\bibitem{CMB-a}
Y. Ali-Ha{\"i}moud and M. Kamionkowski, Phys. Rev. D \textbf{95}, 043534 (2017).
%
\bibitem{CMB-b}
V. Poulin, P. D. Serpico, F. Calore, S. Clesse, and K. Kohri, Phys. Rev. D {\bf 96}, 083524 (2017).
%
\bibitem{EROS}
 P. Tisserand et al. (EROS-2 Collaboration), Astron. Astrophys. {\bf 469}, 387 (2007).
%
\bibitem{kepler}
 K. Griest, A. M. Cieplak, and M. J. Lehner, Phys. Rev. Lett. {\bf 111}, 181302 (2013).
%
\bibitem{HSC}
H. Niikura et al., Nat. Astron. {\bf 3}, 524 (2019).
%
\bibitem{WD}
P. W. Graham, S. Rajendran, and J. Varela, Phys. Rev. D {\bf 92}, 063007 (2015).
%


\bibitem{femto}
A. Barnacka, J. F. Glicenstein, and R. Moderski, Phys. Rev. D {\bf 86}, 043001 (2012).
%
\bibitem{EGG}
B. J. Carr, K. Kohri, Y. Sendouda, and J. Yokoyama, Phys. Rev. D {\bf 81}, 104019 (2010)
%


\bibitem{Katz:2018}
A. Katz, J. Kopp, S. Sibiryakov, and W. Xue, JCAP {\bf 12}, 005 (2018).
%
\bibitem{Montero:2019}
P. Montero-Camacho, X. Fang, G. Vasquez, M. Silva, and C. M. Hirata, JCAP {\bf 08}, 031 (2019).
%
\bibitem{sato:2019}
G. Sato-Polito, E. D. Kovetz, and M. Kamionkowski, Phys. Rev. D {\bf 100}, 063521 (2019).
%
\bibitem{akrami:2018}
Y. Akrami et al. (Planck Collaboration), A\&A {\bf 641}, A10 (2020).
%
\bibitem{Motohashi:2017}
H. Motohashi and W. Hu, Phys. Rev. D {\bf 96}, 063503 (2017).
%
\bibitem{Passaglia:2019}
S. Passaglia, W. Hu, and H. Motohashi, Phys. Rev. D {\bf 99}, 043536 (2019)
%
\bibitem{Khlopov:2010}
M. Yu. Khlopov, Res. Astron. Astrophys. {\bf 10}, 495 (2010).
%
\bibitem{Belotsky1:2014}
K. M. Belotsky et al, Mod. Phys. Lett. A, {\bf  29}, 1440005 (2014).
%
\bibitem{Germani:2017}
C. Germani and T. Prokopec, Phys. Dark Univ. {\bf 18}, 6 (2017).
%
\bibitem{Di:2018}
H. Di and Y. Gong, JCAP {\bf 07}, 007 (2018).
%
\bibitem{Ezquiaga:2018}
J. M. Ezquiaga, J. Garc´ıa-Bellido, and E. R. Morales, Phys. Lett. B {\bf 776}, 345 (2018).
%
\bibitem{Inomata:2018}
K. Inomata, M. Kawasaki, K. Mukaida, and T. T. Yanagida, Phys. Rev. D {\bf  97}, 043514 (2018).
%
\bibitem{Ballesteros:2018}
G. Ballesteros and M. Taoso, Phys. Rev. D {\bf 97}, 023501 (2018).
%
\bibitem{Ozsoy:2018}
O. Ozsoy, S. Parameswaran, G. Tasinato, and I. Zavala, JCAP {\bf 07}, 005 (2018).
%
\bibitem{shiPi:2018}
S. Pi, Y. l. Zhang, Q. G. Huang and M. Sasaki, JCAP {\bf 05}, 042 (2018).
%
\bibitem{Cai:2018}
Y. F. Cai, X. Tong, D. G. Wang, and S. F. Yan, Phys. Rev. Lett. {\bf 121}, 081306 (2018).
%
\bibitem{chen:2019}
C. Chen and Y. F. Cai, JCAP {\bf 10}, 068 (2019).
%
\bibitem{Ballesteros:2019}
G. Ballesteros, J. B. Jim´enez, and M. Pieroni, JCAP {\bf 06}, 016 (2019).
%
\bibitem{Kamenshchik:2019}
A. Y. Kamenshchik, A. Tronconi, T. Vardanyan, and G. Venturi, Phys. Lett. B {\bf 791}, 201
(2019).
%
\bibitem{Atal:2019}
V. Atal, J. Garriga, and A. Marcos-Caballero, JCAP {\bf 09} 073 (2019).
%
\bibitem{Belotsky:2019}
K. M. Belotsky et al, Eur. Phys. J. C, {\bf  79}, 246 (2019).
%
\bibitem{Dalianis:2019}
I. Dalianis, A. Kehagias, and G. Tringas, JCAP {\bf 01}, 037 (2019).
%
\bibitem{fu:2019}
 C. Fu, P. Wu, and H. Yu, Phys. Rev. D {\bf 100}, 063532 (2019).
%
\bibitem{mishra:2020}
S. S. Mishra and V. Sahni, JCAP {\bf 04},  007 (2020).
%
\bibitem{lin:2020}
J. Lin, Q. Gao, Y. Gong, Y. Lu, C. Zhang, and F. Zhang, Phys. Rev. D {\bf 101}, 103515 (2020).
%
\bibitem{Ballesteros:2020a}
 G. Ballesteros, J. Rey, M. Taoso, A. Urbano, JCAP {\bf 07}, 025 (2020).
%
\bibitem{Ballesteros:2020b}
G. Ballesteros, J. Rey, F. Rompineve, JCAP {\bf 06}, 014 (2020).
%
\bibitem{Braglia:2020}
M. Braglia, D. K. Hazrad, F. Finelli, G. F. Smoot, L. Sriramkumari and A. A. Starobinsky, JCAP {\bf 08}, 001 (2020).
%
\bibitem{Fumagalli:2020a}
J. Fumagalli, S. Renaux-Petel, J. W. Ronayne, and L. T. Witkowski,  arXiv:2004.08369.
%
\bibitem{Sypsas:2020}
G. A. Palma, S. Sypsas, and C. Zenteno, Phys. Rev. Lett. {\bf 125}, 121301 (2020).
%
\bibitem{Braglia2:2020}
M. Braglia, X. Chen, and D. K. Hazra, JCAP {\bf 03}, 005 (2021).
%
\bibitem{Dalianis:2020}
 I. Dalianis, S. Karydas, and E. Papantonopoulos, JCAP {\bf 06}, 040 (2020).
%
\bibitem{Sasaki:2018}
 M. Sasaki, T. Suyama, T. Tanaka, and S. Yokoyama, Class. Quant. Grav. {\bf 35}, 063001 (2018).
%
\bibitem{Matarrese:1998}
 S. Matarrese, S. Mollerach, and M. Bruni, Phys. Rev. D {\bf 58}, 043504 (1998).
%
\bibitem{Mollerach:2004}
 S. Mollerach, D. Harari, and S. Matarrese, Phys. Rev. D {\bf 69}, 063002 (2004).
%
\bibitem{Saito:2009}
 R. Saito and J. Yokoyama, Phys. Rev. Lett. {\bf 102}, 161101 (2009.)
%
\bibitem{Garcia:2017}
 J. Garcia-Bellido, M. Peloso, and C. Unal, JCAP {\bf 1709}, 013 (2017).
%
\bibitem{Kohri:2018}
 K. Kohri and T. Terada, Phys. Rev. D {\bf 97}, 123532 (2018).
 %
\bibitem{Cai:2019-a}
R. G. Cai, S. Pi, and M. Sasaki, Phys. Rev. Lett. {\bf 122}, 201101 (2019).
%
\bibitem{Cai:2019-b}
R. G. Cai, S. Pi, S. J. Wang, and X. Y. Yang, JCAP {\bf 05}, 013 (2019).
%
\bibitem{Cai:2019-c}
Y. F. Cai, C. Chen, X. Tong, D. G. Wang, and S. F. Yan, Phys. Rev. D {\bf 100}, 043518 (2019).
%
\bibitem{Bartolo:2019-a}
N. Bartolo, V. De Luca, G. Franciolini, A. Lewis, M. Peloso, and A. Riotto, Phys. Rev. Lett.
{\bf 122}, 211301 (2019).
%
\bibitem{Bartolo:2019-b}
 N. Bartolo, V. De Luca, G. Franciolini, M. Peloso, D. Racco, and A. Riotto, Phys. Rev. D
{\bf 99}, 103521 (2019).
%
\bibitem{Wang:2019}
S. Wang, T. Terada, and K. Kohri, Phys. Rev. D {\bf 99}, 103531 (2019).
%
\bibitem{Fumagalli:2020b}
J. Fumagalli, S. Renaux-Petel, and L. T. Witkowski, JCAP {\bf 08}, 030  (2021).
%
\bibitem{Hajkarim:2019}
F. Hajkarim and J. Schaffner-Bielich, Phys. Rev. D {\bf 101}, 043522 (2020).
%
\bibitem{Xu:2020}
W. T. Xu, J. Liu, T. J. Gao, and Z. K. Guo, Phys. Rev. D {\bf 101}, 023505 (2020).
%
\bibitem{Domenech:2020a}
G. Dom\`enech, and M. Sasaki, Int. J. Mod. Phys. D {\bf 29}, 2050028 (2020).
%
\bibitem{Domenech:2020b}
G. Dom\`enech, and M. Sasaki, Phys. Rev. D {\bf 103}, 063531 (2021).
%
\bibitem{Horndeski:1974}
G. W. Horndeski, Int. J. Theor. Phys. {\bf 10}, 363 (1974).
%
\bibitem{kobayashi:2010}
 T. Kobayashi, M. Yamaguchi, J. Yokoyama, Phys. Rev. Lett. {\bf 105}, 231302 (2010).
%
\bibitem{Burrage:2010}
C. Burrage, C. d. Rham, D. Seery, and A. J. Tolley, JCAP {\bf 01}, 014 (2011).
%
\bibitem{teimoori:2018}
 Z. Teimoori and K. Karami, Astrophys. J. {\bf 864}, 41 (2018).
%
\bibitem{Tumurtushaa:2019}
G. Tumurtushaa,  Eur. Phys. J. C {\bf 79}, 920 (2019).
%
\bibitem{Nicolis:2009}
A. Nicolis, R. Rattazzi, E. Trincherini, Phys. Rev. D {\bf 79}, 064036 (2009).
%
\bibitem{Deffayet:2009a}
 C. Deffayet, G. Esposito-Farese, A. Vikman, Phys. Rev. D {\bf 79}, 084003 (2009).
%
\bibitem{Deffayet:2009b}
C. Deffayet, S. Deser, G. Esposito-Farese, Phys. Rev. D {\bf 80}, 064015 (2009)
%
\bibitem{Hirano:2016}
S. Hirano, T. Kobayashi and S. Yokoyama, Phys. Rev. D {\bf 94}, 103515 (2016).
%
\bibitem{Ohashi:2012}
J. Ohashi and S. Tsujikawa, JCAP {\bf 10}, 035 (2012).
%

\bibitem{Fixsen:1996}
 D. J. Fixsen, E. S. Cheng, J. M. Gales, J. C. Mather, R. A. Shafer, and E. L. Wright, Astrophys. J. {\bf 473}, 576 (1996).
%
\bibitem{Inomata:2016}
K. Inomata, M. Kawasaki, and Y. Tada, Phys. Rev. D {\bf 94}, 043527 (2016).
%
\bibitem{Inomata:2019-a}
 K. Inomata and T. Nakama, Phys. Rev. D {\bf 99}, 043511 (2019).
%
\bibitem{mahbub:2020}
R. Mahbub, Phys. Rev. D {\bf 101}, 023533 (2020).
%
%
\bibitem{Shibata:1999}
M. Shibata and M. Sasaki, Phys. Rev. D {\bf 60} 084002 (1999).
%
\bibitem{Polnarev:2007}
A. G. Polnarev and I. Musco, Class. Quant. Grav. {\bf 24}, 1405 (2007).
%
\bibitem{Musco:2009}
I. Musco, J. C. Miller and A. G. Polnarev, Class. Quant. Grav. {\bf 26}, 235001 (2009).
%
\bibitem{harada:2013}
 T. Harada, C.-M. Yoo, and K. Kohri, Phys. Rev. D {\bf 88}, 084051 (2013).
%
\bibitem{Musco:2013}
I. Musco and J. C. Miller, Class. Quant. Grav. {\bf 30}, 145009 (2013).
%
\bibitem{young:2014}
S. Young, C. T. Byrnes, and M. Sasaki, JCAP {\bf 07}, 045 (2014).
%
\bibitem{Germani:2019}
C. Germani and I. Musco, Phys. Rev. Lett. {\bf 122}, 141302 (2019)
%
\bibitem{lisa}
P. Amaro-Seoane et al. (LISA Collaboration), arXiv:1702.00786.
%
\bibitem{Ananda:2007}
 K. N. Ananda, C. Clarkson, and D. Wands, Phys. Rev. D {\bf 75}, 123518 (2007).
%
\bibitem{Baumann:2007}
 D. Baumann, P. J. Steinhardt, K. Takahashi, and K. Ichiki, Phys. Rev. D {\bf 76}, 084019 (2007).
%
\bibitem{EPTA-a}
R. D. Ferdman et al., Class. Quant. Grav. {\bf 27}, 084014 (2010).
%
\bibitem{EPTA-b}
G. Hobbs et al., Class. Quant. Grav. {\bf 27}, 084013 (2010).
%
\bibitem{EPTA-c}
M. A. McLaughlin, Class. Quant. Grav. {\bf 30}, 224008 (2013).
%
\bibitem{EPTA-d}
G. Hobbs, Class. Quant. Grav. {\bf 30}, 224007 (2013).
%
\bibitem{ska}
C. J. Moore, R. H. Cole, and C. P. L. Berry, Class. Quant. Grav. {\bf 32}, 015014 (2015).
%
\bibitem{ligo-a}
G. M. Harry (LIGO Scientific Collaboration), Class. Quant. Grav. {\bf 27}, 084006 (2010).
%
\bibitem{ligo-b}
J. Aasi et al. (LIGO Scientific Collaboration), Class. Quant. Grav. {\bf 3}2, 074001 (2015).
%
\bibitem{lisa-a}
K. Danzmann, Class. Quant. Grav. {\bf 14}, 1399 (1997).
%
\bibitem{taiji}
W. R. Hu and Y. L. Wu, Natl. Sci. Rev. {\bf 4}, 685 (2017).
%
\bibitem{tianqin}
J. Luo et al. (TianQin Collaboration), Class. Quant. Grav. {\bf 33}, 035010 (2016).
%
\bibitem{Fu:2020}
C. Fu, P. Wu, and H. Yu, Phys. Rev. D {\bf 101}, 023529 (2020).
%
\bibitem{Kuro:2018}
S. Kuroyanagi, T. Chiba, and T. Takahashi, JCAP {\bf 11}, 038 (2018).
%
\bibitem{Yuan:2020}
C. Yuan, Z. C. Chen, and Q. G. Huang, Phys. Rev. D {\bf 101}, 043019 (2020).
%
\bibitem{shipi:2020}
R. G. Cai, S. Pi, and M. Sasaki, Phys. Rev. D {\bf 102}, 083528 (2020).
%
\end{thebibliography}
\end{document}